\documentclass[utf8]{frontiersinFPHY_FAMS} 

\setcitestyle{square} 

\usepackage[onehalfspacing]{setspace}
\usepackage{url,hyperref,lineno,microtype,subcaption,datetime,fmtcount,etoolbox,fcprefix, amsmath, amsthm, amsfonts}
\usepackage[super]{nth}
\usepackage[upgreek]{mathastext}

\usepackage{revsymb}
\usepackage{orcidlink}

\def\keyFont{\fontsize{8}{11}\helveticabold }
\def\firstAuthorLast{Barty {et~al.}} 
\def\Authors{Christopher P. J. Barty\,\orcidlink{0000-0001-8019-5976}$^{1,2,3,*}$,
J. Martin Algots\,$^{1}$,
Alexander J. Amador\,$^{1}$,
James C. R. Barty\,$^{1}$,
Shawn M. Betts\,$^{1}$,
Marcelo A. Castañeda\,$^{1}$,
Matthew M. Chu\,$^{1}$,
Michael E. Daley\,$^{1}$,
Ricardo A. De Luna Lopez\,$^{1}$,
Derek A. Diviak\,$^{1}$,
Haytham H. Effarah\,\orcidlink{0000-0002-4483-6745}$^{1,2,3}$,
Roberto Feliciano\,$^{1}$,
Adan Garcia\,$^{1}$,
Keith J. Grabiel\,$^{1}$,
Alex S. Griffin\,$^{1}$,
Frederic V. Hartemann\,$^{1}$,
Leslie Heid\,$^{1,2}$,
Yoonwoo Hwang\,\orcidlink{0000-0003-2162-8094}$^{1}$,
Gennady Imeshev\,$^{1}$,
Michael Jentschel\,\orcidlink{0000-0001-8932-4682}$^{1}$,
Christopher A. Johnson\,$^{1}$,
Kenneth W. Kinosian\,$^{1}$,
Agnese Lagzda\,\orcidlink{0000-0002-5672-2112}$^{1}$,
Russell J. Lochrie\,$^{1}$,
Michael W. May\,$^{1}$,
Everardo Molina\,$^{1}$,
Christopher L. Nagel\,$^{1}$,
Henry J. Nagel\,$^{1}$,
Kyle R. Peirce\,$^{1}$,
Zachary R. Peirce\,$^{1}$,
Mauricio E. Quiñonez\,$^{1}$,
Ferenc Raksi\,$^{1}$,
Kelanu Ranganath\,$^{1}$,
Trevor Reutershan\,\orcidlink{0000-0001-7740-5741}$^{1,2,3}$,
Jimmie Salazar\,$^{1}$,
Mitchell E. Schneider\,\orcidlink{0000-0002-5996-0431}$^{1}$,
Michael W. L. Seggebruch\,\orcidlink{0000-0002-2813-3003}$^{1,2}$,
Joy Y. Yang\,$^{1}$,
Nathan H. Yeung\,$^{1}$,
Collette B. Zapata\,$^{1}$,
Luis E. Zapata\,\orcidlink{0000-0002-3760-4145}$^{1}$,
Eric J. Zepeda\,$^{1}$,
Jingyuan Zhang\,\orcidlink{0000-0002-9620-503X}$^{1}$
}

\def\Address{$^{1}$Lumitron Technologies, Inc., Irvine, CA, United States \\
$^{2}$Physics and Astronomy Department, University of California, Irvine, CA, United States \\ 
$^{3}$Beckman Laser Institute and Medical Clinic, University of California, Irvine, CA, United States}
\def\corrAuthor{Christopher P. J. Barty}

\def\corrEmail{cbarty@uci.edu}

\begin{document}
\onecolumn
\firstpage{1}

\title[Distributed Charge Compton Source]{Design, Construction, and Test of Compact, Distributed-Charge, X-Band Accelerator Systems that Enable Image-Guided, VHEE FLASH Radiotherapy} 

\author[\firstAuthorLast ]{\Authors} 
\address{} 
\correspondance{} 

\extraAuth{}

\maketitle

\begin{abstract}

\section{}
The design and optimization of laser-Compton x-ray systems based on compact distributed charge accelerator structures can enable micron-scale imaging of disease and the concomitant production of beams of Very High Energy Electrons (VHEEs) capable of producing FLASH-relevant dose rates.
The physics of laser-Compton x-ray scattering ensures that the scattered x-rays follow exactly the trajectory of the incident electrons, thus providing a route to image-guided, VHEE FLASH radiotherapy.
The keys to a compact architecture capable of producing both laser-Compton x-rays and VHEEs are the use of X-band RF accelerator structures which have been demonstrated to operate with over 100 MeV/m acceleration gradients.
The operation of these structures in a distributed charge mode in which each radiofrequency (RF) cycle of the drive RF pulse is filled with a low-charge, high-brightness electron bunch is enabled by the illumination of a high-brightness photogun with a train of UV laser pulses synchronized to the frequency of the underlying accelerator system.
The UV pulse trains are created by a patented pulse synthesis approach which utilizes the RF clock of the accelerator to phase and amplitude modulate a narrow band continuous wave (CW) seed laser.
In this way it is possible to produce up to 10 \textmu A of average beam current from the accelerator. Such high current from a compact accelerator enables production of sufficient x-rays via laser-Compton scattering for clinical imaging and does so from a machine of "clinical" footprint.
At the same time, the production of 1000 or greater individual micro-bunches per RF pulse enables $>$~10~nC of charge to be produced in a macrobunch of $<$~100~ns. The design, construction, and test of the 100-MeV class prototype system in Irvine, CA is also presented.

\tiny
 \keyFont{ \section{Keywords:} Lasers, x-rays, laser-Compton scattering, accelerators, x-band, FLASH, high-resolution radiography, VHEE } 
\end{abstract}

\section{Introduction}

The Distributed Charge Compton Source (DCCS) \cite{barty-patent} architecture and its underlying electron accelerator system are a solution for compact, image-guided, ultra-high dose rate (UHDR), very high energy electron (VHEE) radiation therapy systems.
VHEEs (electron energy $>$ 50 MeV) have been identified as a promising ionizing radiation modality, but the current clinical applicability of VHEE technology is limited \cite{ronga-vhee-review, vozenin-clinical-FLASH}.
An ideal clinical VHEE source would be compact, capable of UHDR operation to potentially leverage the FLASH effect \cite{favaudon:2014} (dose-rate dependent sparing of healthy tissue with dose-rate independent tumor kill), and administered with reliable image guidance \cite{image-guidance-FLASH, taylor:2022, garibaldi:2024}.
An ideal clinical x-ray imaging source based on laser-Compton scattering (LCS), sometimes known as inverse Compton scattering (ICS) \cite{gunther:ics-review}, would also be compact, have a micron-scale effective source size for high-resolution imaging, and able to produce sufficient x-ray flux for clinically relevant phase contrast and/or spectral contrast imaging.
We argue here that the linear electron accelerator (LINAC) required for these two applications is optimized by a distributed charge architecture.

\begin{figure}[ht!]
\begin{center}
\includegraphics[width=17.5cm]{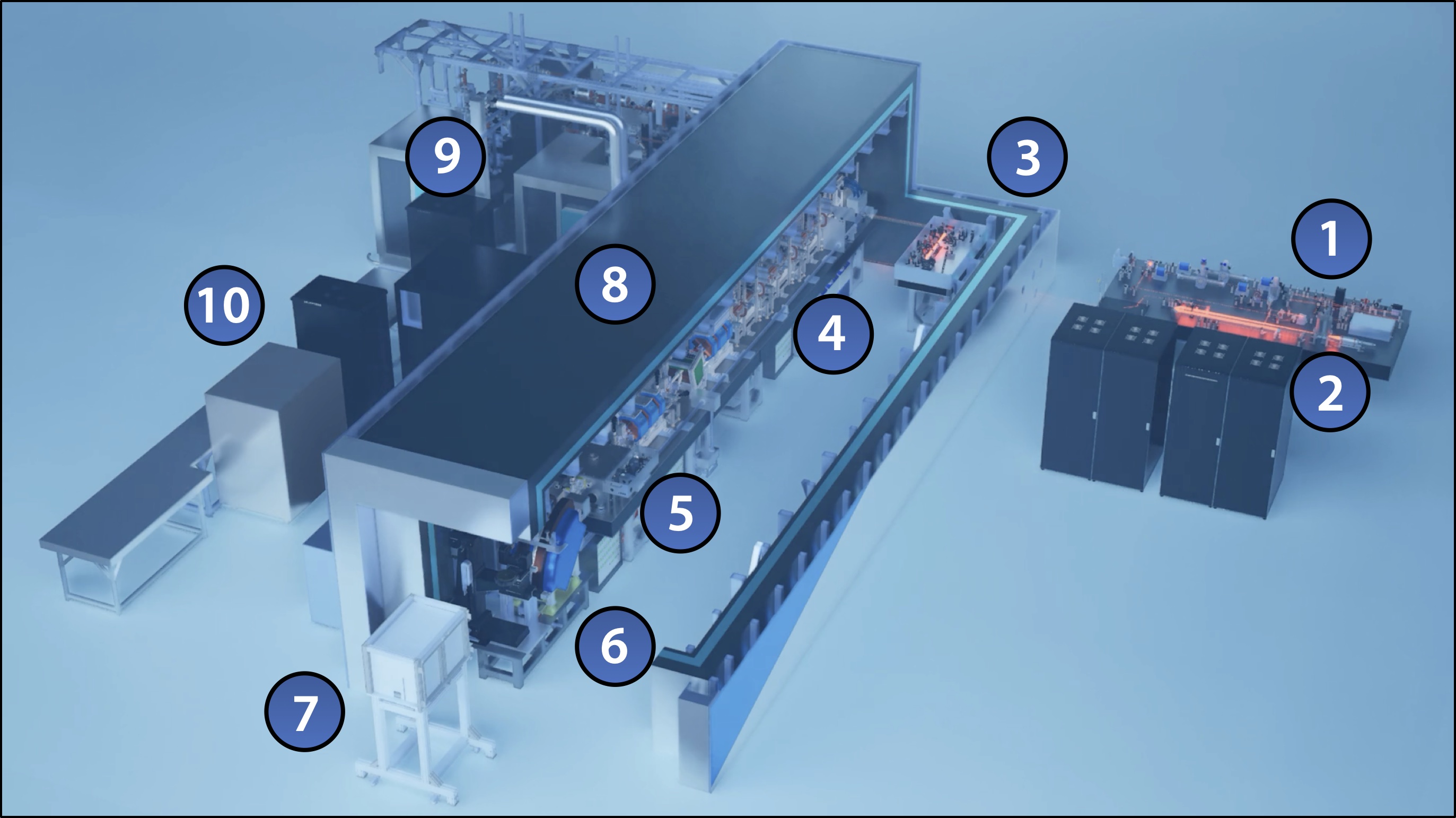}
\end{center}
\caption{CAD model of the Distributed Charge Compton Source (DCCS) at Lumitron Technologies, Inc. in Irvine, CA. (1) Multi-GHz Interaction Laser System (ILS), (2) multi-GHz Photogun Laser System (PGL), (3) multi-pass cell, (4) X-band accelerator beamline, (5) laser-Compton interaction chamber, (6) electron beam dump, (7) X-ray and $\gamma$-ray imaging systems, (8) 9.6-m long custom radiation enclosure, (9) X-band radiofrequency power systems, (10) control station. For a detailed animated fly-through of the DCCS electron acceleration and laser-electron interaction process, see Ref. \cite{yang_barty_2024}.}
\label{fig:DCCS-CAD}
\end{figure}

\begin{figure}[ht!]
\begin{center}
\includegraphics[width=17.5cm]{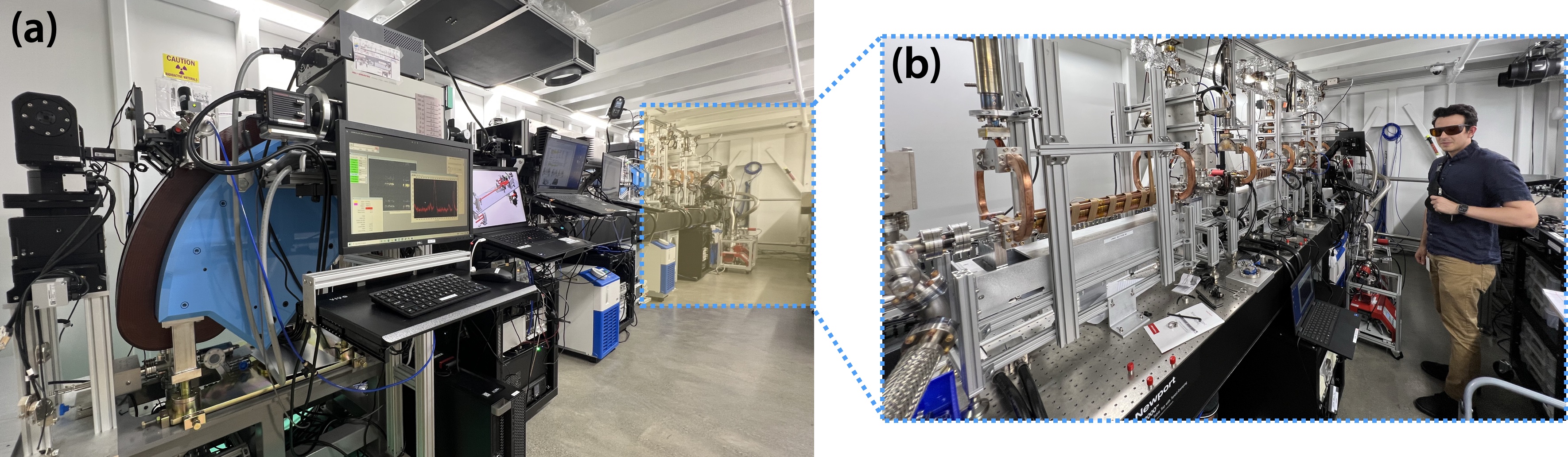}
\end{center}
\caption{(a) Photograph of the  Distributed Charge Compton Source (DCCS) at Lumitron Technologies, Inc. in Irvine CA inside the radiation safety enclosure (bunker). The total length of the bunker is 9.6 m. Highlighted in yellow with a dotted outline is the portion of the beamline dedicated to electron acceleration. (b) A close up of the installed and operational DCCS accelerator. The total length of accelerator sections capable of producing VHEEs is approximately 3~m.}\label{fig:DCCS-pic}
\end{figure}

``Distributed charge'' is a strategy to increase the average current of a linear accelerator by distributing electrons over many bunches separated by a single radiofrequency (RF) period instead of maximizing the number of electrons in a single bunch.
In compact RF accelerators with high operation frequencies into the X band (8 - 12 GHz), the total number of electrons that can be effectively accelerated in a single bunch decreases, especially when trying to preserve electron beam quality \cite{rosenzweig:1995}.
A distributed charge architecture allows the production of enough electrons for both LCS and VHEE applications while maintaining a compact accelerator footprint.
Figure \ref{fig:DCCS-CAD} is a CAD model of the currently operational DCCS at Lumitron Technologies, Inc. with primary systems labeled.
Figure \ref{fig:DCCS-pic}(a) is a photograph of inside of the radiation safety enclosure of the compact MeV-class laser-Compton light source and VHEE system at Lumitron Technologies, Inc. in Irvine, CA.
Figure \ref{fig:DCCS-pic}(b) is a close-up of the compact VHEE accelerator of the DCCS at Lumitron Technologies (item 4 in Figure \ref{fig:DCCS-CAD}).
This accelerator is designed to produce electrons with energies up to 100 MeV with sufficient charge for UHDR operation.
For a detailed animated fly-through of the DCCS electron acceleration and laser-electron interaction process, see Ref. \cite{yang_barty_2024}.

In this work, we describe the distributed charge architecture and discuss its advantages as a laser-Compton x-ray source for clinically relevant x rays capable of high-resolution, narrow-bandwidth imaging and its advantages as a radiotherapy source of VHEEs for applications in clinically relevant UHDR operations.
The results of systems integration tests of the DCCS in Irvine, CA with respect to production of both x rays and FLASH-relevant electron beams are presented.
To conclude, the potential for the DCCS architecture to serve as a framework for x-ray image-guided VHEE FLASH radiotherapy is discussed.

\section{Motivation}
\label{sec:motivation}

\subsection{Laser-Compton X-Ray Sources}
\label{sec:compton}

Compact laser-Compton light sources are capable of creating quasi-monoenergetic, tunable, x rays and $\gamma$ rays with outputs that are similar to those of much larger, km-scale, international synchrotron facilities \cite{hartemann:2005:prab, jacquet:2016}.
Synchrotron facilities have demonstrated the potential for clinically-relevant imaging techniques that leverage narrow spectral bandwidths for K-edge subtraction imaging \cite{thomlinson:2018} and small effective x-ray source sizes for phase contrast imaging \cite{liu:2013, hip-ct}.
Although synchrotrons are capable of imaging modalities currently unfeasible for conventional x-ray tube technologies, the cost and size of synchrotron facilities is prohibitive for widespread clinical use.
Since the first conceptions of a laser-Compton x-ray source in 1963 \cite{milburn:1963, arutyunian:1963}, just three years after the first experimental demonstration of the laser \cite{maiman:1960}, and the first demonstrations of laser-Compton scattering \cite{fiocco:1963, kulikov:1964, bemporad:1965}, a rich field has developed with many differing strategies for optimizing laser-Compton x-ray sources for various applications and different energy regimes \cite{gunther:ics-review}.

Briefly, the laser-Compton interaction can be described as the interaction of short-duration, energetic laser pulses with bright, monoenergetic, relativistic electrons. This interaction induces a transverse motion on the electron bunch, which in turn radiates as an electric dipole and produces Doppler-upshifted x rays and/or $\gamma$ rays in the laboratory frame of reference.
In effect, the laser-Compton interaction performs a similar function to that of the periodic array of magnets of a synchrotron wiggler \cite{brown:1983}.
In the laboratory frame, the Compton “scattered” photons appear to be higher energy than incident photons and as such this process is sometimes referred to as “inverse” Compton scattering \cite{jones:1965}. The laboratory-frame formula to describe the energy of Compton-scattered x rays in a head-on collision of counter-propagating photons and electrons can be described as
\begin{equation}
    E_{ph} = \frac{4 \gamma^2}{1 + \gamma^2 \theta^2 + 4 \gamma k_0 {\lambdabar}} E_L,
\label{eq:1}
\end{equation}
where $E_{ph}$ is the scattered-photon energy, $E_L$ is the incident laser photon energy, $\gamma$ is the electron relativistic factor, $\theta$ is a small observation angle relative to the electron propagation axis, $k_0$ is the incident laser photon angular wavenumber, and $\lambdabar$ is the reduced Compton wavelength of the electron.
However, in the rest frame of the moving electron, the Doppler effect leads to the incident photon appearing to be higher frequency. In the rest frame of the electron, the scattered photon is lower frequency, just as in traditional Compton scattering \cite{compton:1923}.
To remove any confusion, we refer to the relativistic x-ray generation process as laser-Compton scattering. 

The primary advantage of using a laser's electric field to induce periodic motion in relativistic electrons is that the induced oscillations can occur at much higher spatial frequencies than what are achievable by a periodic magnet array.
This rapid laser-induced oscillation drastically reduces the energy requirements of the interacting electron beam, which has enabled the production of x rays and $\gamma$ rays from 6~keV to 1~GeV using many different architectural designs \cite{gunther:ics-review, gibson:2004, weller:2009, barty:2004:trex}.

The primary limitation of laser-Compton sources is the small Thomson cross section ($6 \times 10^{-25} \; \text{cm}^2$) for laser-scattering from the relativistic electrons.
To overcome this limitation, the most efficient laser-Compton systems operate in a co-focused geometry in which both the electron bunch and laser pulses are brought to a common focus and are synchronized so that both entities arrive at that focus at the same time.
While laser-Compton scattering can be configured for any angle of incidence between the electron and the laser pulses, the head-on or 180-degree configuration yields the highest on-axis Doppler upshift and provides the most tolerance with respect to errors in arrival timing between the electron bunch and laser pulse.
Ideally, the duration of the laser pulses and electron bunches are both on the order of the transit time of the focal region or less.
Tuning of the x-ray energy can be accomplished by changing the energy of the laser photons \cite{drebot:2017}, changing the laser-electron interaction angle \cite{suerra:2021}, or changing the energy of the electron bunch.
In the DCCS architecture described here, changing the energy of the electron bunch is the most practical approach.

By conservation of energy and momentum, the spectrum of a laser-Compton source is angle-correlated (Eq \ref{eq:1}).
The spectrum of a 180-degree incidence configuration ranges from about $4 \gamma^2 E_L$ in the direction of the electron bunch to half that value for photons scattered at 90$^{\circ}$ relative to the electron trajectory.
By placing an aperture in the generated beam path, the integrated bandwidth of the transmitted beam may be reduced until such point that the energy and angle variations of the electrons and photons involved in the Compton scattering process dominate the effect on the bandwidth.
At this point, reducing the aperture size simply reduces the total flux without changing the bandwidth.
Laser-Compton sources have typically achieved on-axis bandwidths of between 3\% and 12\% \cite{schoenlein, hartemann:2004:pleiades-characterization, weller:2009, albert:2010:PRSTAB, gibson:2010:PRSTAB, jacquet:2024:thomx-first-light, tilton:2024, sakai:2024}.
However, an optimized laser-Compton source based on high-brightness, monoenergetic electron bunches and high-beam-quality, picosecond laser pulses, can theoretically produce a minimum, on-axis bandwidth of 0.1\% full width at half maximum (FWHM).
The first experimental confirmation of a laser-Compton source design with exceptionally narrow on-axis x-ray bandwidth was demonstrated using the compact Energy Recovery Linac (cERL) at KEK with a measured on-axis x-ray bandwidth of 0.4\% operating at 6.95 keV \cite{akagi:2016}.
The cERL approach to producing high-brightness, monoenergetic electron bunches was to use superconducting accelerators.
While this approach provides small on-axis energy bandwidths with demonstrated imaging capabilities \cite{Kosuge:2018}, albeit at sub-clinical x-ray energies ($<$ 20~keV), the widespread adoption of superconducting accelerator architectures is prohibited by the cost, size, and complexity of the required infrastructure.

There are two generic approaches to production of clinically-relevant x rays via laser-Compton scattering. These approaches can be separated by the underlying accelerator architecture which is either that of an electron storage ring or that of a linear accelerator (LINAC). 

\textbf{Storage ring laser-Compton systems.} In the storage ring approach, an energetic electron bunch is injected into a closed-loop magnetic lattice and “stored” for a number of round trips.
On each round trip the bunch interacts with a synchronized laser pulse to produce Compton x rays.
Due to synchrotron losses and imperfections in the lattice, the electron bunch quality will decay over time. 
At some point, the circulating electrons are ejected from the cavity and a new bunch is injected. 
The advantage of a storage ring approach is that the electron beam average current can be high thus, in principle, increasing the laser-Compton x-ray flux from the machine.
The disadvantages of storage ring approaches are;
a) the electron bunches cannot be focused to a tight spot without destroying the electron beam quality which limits the potential output flux,
b) the laser average power required to achieve high flux cannot be readily obtained without the use highly-sensitive, resonant enhancement cavities, 
c) the electron beam characteristics (emittance, energy, and energy spread) change over the course of its lifetime within the storage ring, 
d) the synchronization of the laser and electron bunch timing at the interaction region is non-trivial,
e) the tuning of the x-ray energy via changes in the electron beam energy is limited by the speed with which modifications of the magnetic lattice and the electron beam injector can be made, and
f) the charge of the stored electron bunch is not sufficient to be of practical use as an UHDR clinical electron irradiation source.

Nonetheless, storage ring-based laser-Compton systems have produced beams with clinically relevant x-ray energy (MuCLS \cite{eggl:2016} and ThomX \cite{jacquet:2016}) and above (HI$\gamma$S \cite{weller:2009}).
Pre-clinical studies at the Munich Compact Light Source (MuCLS) have demonstrated spectral contrast imaging and phase contrast imaging applications that are much more difficult or impossible to perform with conventional x-ray tube sources with reported x-ray fluxes up to $4.5 \times 10^{10}$ photons per second at energies up to 35~keV \cite{gunther:2020}.
The main limitations of the MuCLS are its upper energy limit (electrons: 45 MeV; x rays: 35 keV) and its 50-\textmu m x-ray root-mean-square (RMS) source radius.
The ThomX collaboration, which recently demonstrated first x-ray production \cite{jacquet:2024:thomx-first-light}, also uses a storage ring architecture.
Although improvements in their system are expected to increase x-ray energy to up to 90 keV, their minimum effective x-ray source size is currently no less than 65 \textmu m RMS based on their interaction laser spot size.

\textbf{LINAC laser-Compton systems.} In the LINAC based approaches to laser-Compton sources, a new electron bunch is generated from a laser-driven photogun for each laser-Compton interaction, i.e. the electrons are used once and discarded. Doing so allows the electrons be focused to much smaller spot sizes than possible in storage ring based systems.  This increases the output flux of the system per electron and enhances the imaging capabilities of the device.  Tuning of the system from one x-ray energy to another can be done rapidly (seconds to minutes) by changing the RF power to the accelerator sections of the LINAC.  The electrons produced by the system can have sufficient charge and pulse structure to enable VHEE and VHEE FLASH radiotherapy.  Traditionally, LINAC-based systems have operated with one electron bunch per RF pulse driving the accelerator. The downsides of this LINAC based approach are:
a) average beam current is limited by the repetition rate of the RF power system and the charge that may be stably accelerated by the system, 
b) the photogun laser system which produces the initial electrons must be timed precisely with respect to the phase of the RF driving the accelerator, and
c) the interaction laser system which creates the laser pulses that interact with the focused electron bunches to produce Compton x rays must be timed precisely with respect to the electrons.
The above disadvantages are eliminated via a distributed charge Compton source architecture \cite{barty-patent} that uses RF pulse synthesis to create the photogun and interaction laser pulse trains. 

The exceptional electron beam quality afforded by the DCCS architecture enables the production of an x-ray beam with an RMS source radius below 5 \textmu m, and a total x-ray flux greater than $10^{12}$ photons/second (see Section \ref{sec:modeling}).
The current upper energy limits of the DCCS also expand the flexibility of its applications with electron energies up to 100 MeV and x-ray energies up to 360 keV, enabling both the investigation of VHEE irradiation and nuclear-based x-ray imaging techniques.

Using compact, normal-conducting, RF accelerator technology is necessary for the eventual clinical translation of laser-Compton technology.
To produce x rays with the most exquisite quality to leverage the spectral and phase-based imaging modalities developed at synchrotron facilities, high-brightness electron beams must be used to minimize the on-axis x-ray bandwidth and effective x-ray source size.
The quality (brightness) of an electron beam is most readily achieved with relatively little charge in each electron bunch, especially in high-frequency RF accelerators (Section \ref{sec:photogun}).
Counteracting this charge limitation to obtain a clinically relevant x-ray flux requires a distribution of charge across long, consecutive trains of electron bunches (Section \ref{sec:pulse-synthesis}).
We argue here that the DCCS architecture is the solution for clinically translatable laser-Compton x-ray sources that enable spectral and phase-based imaging techniques.

\subsection{Very High Energy Electron (VHEE) Sources}
\label{sec:vhee-sources}

There is urgent need for transformative technologies in compact, high-gradient accelerator architectures that enable both VHEE and FLASH capabilities \cite{ronga-vhee-review}.
Compared to photon or proton radiation sources, electron sources are most readily capable of achieving ultra-high dose rates \cite{dunning:2019, giannini:2024} and, in the VHEE regime, will have appropriate penetration to treat deep-seated tumors in humans \cite{desrosiers:2000, sarti:2021, bohlen:2024}.
One of the tightest bottlenecks in investigating the FLASH effect is simply the lack of availability of appropriate ionizing radiation sources \cite{giannini:2024}.
More VHEE sources are being designed and commissioned to address this need \cite{gamba:2018, maxim:2019, stephan:2022, faillace:2022, palumbo:2023, lin:2024} with varying strategies regarding accelerator design.

VHEE research opportunities and clinical translatability are both fundamentally limited by facility size requirements.
The first reported VHEE dosimetry experiments were performed at the Oak Ridge electron linear accelerator (ORELA) \cite{desrosiers:2000}, at the Sources for Plasma Accelerators and Radiation Compton with Lasers and Beams (SPARC) S-band beamline \cite{subiel:2014} and at the Next Linear Collider Test Accelerator (NLCTA) S-band/X-band beamline \cite{bazalova-carter:2015}.
Recent VHEE experiments at the CERN Linear Electron Accelerator Research (CLEAR) facility continue to garner interest in VHEE, especially with potential for UHDR operation, through investigating VHEE dosimetry \cite{poppinga:2020:clear, small:2021:clear}, VHEE insensitivity to tissue inhomogeneity \cite{lagzda:2020:clear}, VHEE beam focusing \cite{kokurewicz:2024:clear-focusing, whitmore:2024:clear-focusing}, and techniques for UHDR VHEE dose monitoring \cite{bateman:2024:clear-dosimeter, hart:2024:clear-dosimetry}.
Even with promising VHEE results from the CLEAR facility, the total beamline length of 41 meters (25~m injector + 16~m beamline, \cite{gamba:2018}) limits the practicality of widespread clinical adoption.
Another VHEE collaboration, the FLASHlab@PITZ, is being commissioned to further investigate VHEE with charge-per-pulse values deep into UHDR regime, but at the cost of a clinically impractical footprint using L-band accelerators (1.4 GHz) \cite{stephan:2022}. The FLASHlab@PITZ requires the existing 22-meter photoinjector to reach 22 MeV and an additional 20 meters of planned accelerators to reach 250 MeV. 
SAFEST, a recently announced collaboration between Sapienza University and Istituto Nazionale di Fisica Nucleare (INFN), is seeking to address the VHEE facility size problem by operating at C-band (5.712 GHz), with an anticipated final beamline length of around 5 meters to reach up to 130 MeV \cite{faillace:2022, palumbo:2023}.
Finally, a research team at Tsinghua University has also proposed a compact VHEE accelerator design to reach up to 100~MeV electrons using X-band (11.424 GHz) accelerators with an anticipated total beamline length of less than 2~meters \cite{lin:2024}.
The proposed Tsinghua approach is similar to that used by Lumitron.
In this regime, the accelerator hardware is small compared to the underlying RF power components and thus is no longer the limiting factor with respect to reducing machine footprint.

While both the SAFEST and Tsinghua designs seek to address the compactness problem for clinical translation of VHEE technology, both designs rely on the use of a high-voltage direct current (DC) thermionic electron gun to produce a large electron current.
This strategy increases the available charge for UHDR operation at the expense of fundamentally limiting the quality of the electron beam.
Comparing the SAFEST and Tsinghua DC gun normalized transverse electron emittance (10~mm-mrad expected, and 7.26 mm-mrad measured, respectively) to the LLNL/SLAC electron gun design (0.3 mm-mrad measured) (See Section \ref{sec:photogun}) emphasizes that the DCCS architecture retains the ability to efficiently produce a high quality diagnostic x-ray beam through laser-Compton scattering while producing sufficient electron current for UHDR VHEE operation.
The DCCS architecture is designed to produce 10 \textmu A of average current when operating at with 1000~microbunches at 400~Hz and at energies up to 100~MeV.
To date, the prototype DCCS accelerator at Lumitron Technologies in Irvine, CA has demonstrated the production of 49~MeV electrons at 14 nC in 86.6 ns at 100~Hz, which corresponds to a an average current of 1.4 \textmu A.
Even though the DCCS accelerator bunch charge is limited by the laser fluence on a photocathode and is thus more challenging to produce current than a DC gun, distributing the charge over long trains of closely spaced bunches overcomes this limitation.
Additionally, variable illumination of the photogun enables precise control of the total electron charge delivered to a patient. 

For these reasons, we posit that the DCCS architecture is not only the solution for clinically relevant laser-Compton x-ray imaging, but also the solution for a clinically-translatable, FLASH-capable, VHEE source. 
The potential to rapidly switch between these modes of operation to enable laser-Compton x-ray image-guided VHEE FLASH radiation therapy will be discussed in Section \ref{sec:image-guided}.

\section{Distributed Charge Compton Source Architecture}
\label{sec:architecture}

The Distributed Charge Compton Source (DCCS) architecture is founded on research and development efforts surrounding laser-Compton scattering (LCS) activities at the Lawrence Livermore National Laboratory (LLNL) \cite{anderson:2004:APB, gibson:2004:PoP, hartemann:2004:LPM, shverdin:2010:OL}, as well as new innovations and systems integration studies that have occurred at Lumitron Technologies, Inc. in Irvine, California.
The DCCS is a patented architecture \cite{barty-patent, messerly-patent} that involves the repeated interaction of trains of electron bunches with trains of laser pulses each of which being spaced at exactly the repetition period of the compact, high-gradient, X-band RF accelerator.
This architecture enables bright electron beams and reduces requirements on the interaction laser.
Extensions of the currently operational DCCS prototype at Lumitron Technologies could produce tunable, 30-keV to 3-MeV, x-ray bursts at an effective 400-kHz repetition rate with a total flux of greater than $10^{12}$ photons/sec and an on-axis bandwidth as low as 0.1\% \cite{reutershan:2022}.

The large x-ray flux and narrow x-ray energy bandwidth of the DCCS are enabled by three core technologies: high-gradient X-band (11.424 GHz) photoguns and LINACs, RF laser-pulse synthesis of 11.424 GHz pulse trains, and diode-pumped infrared (IR) laser technology.

\subsection{Numerical Modeling of the DCCS}
\label{sec:modeling}

In this section, we will describe a representative numerical modeling study that outlines the ideal laser-electron interaction specifications to maximize output flux at $10^{12}$ photons per second with a Distributed Charge Compton source (DCCS) architecture at a fixed interaction laser pulse energy.
A summary of these specifications is presented in Table \ref{table:modeling}.
If one desired to minimize bandwidth, it is possible to do so by increasing the interaction spot size at the expense of output flux, unless the laser pulse energy is increased accordingly.

\begin{table}[ht!]
    \centering
\begin{tabular}{ p{4cm}p{4cm}|p{4cm}p{4cm} }
 \hline
 \multicolumn{2}{c|}{\textbf{Interaction Laser}} &
 \multicolumn{2}{c}{\textbf{Electron Beam}}\\
 \hline
 Wavelength         & 354.67 nm & Beam energy & 137 MeV\\
 Micropulse energy  & 10 mJ     & Bunch charge &   25 pC\\
 $M^2$              & 1         & $\varepsilon_{n, rms}$ &   0.3 mm-mrad\\
 FTL bandwidth  & 93 pm & Energy spread ($\sigma_E / E$)  &   0.05\%\\
 Pulse train length & 100       & Bunch train length & 1000 microbunches\\
 Recirculation efficiency  & 10 & Duration ($\sigma_t$) & 0.59 ps \\
 FWHM duration & 2.0 ps    & Duration ($\sigma_\theta$) &  2.4265$^{\circ}$ \\
 Temporal shape     & Gaussian  & Temporal shape & Gaussian \\
 FWHM diameter & 3.46 \textmu m & Radius ($\sigma_{x,y}$) &  1.5 \textmu m \\
 Focal shape        & Gaussian  & Focal shape & Gaussian \\
 Pulse spacing      & 87.535 ps & X-band frequency & 11.424 GHz \\
 Repetition rate    & 400 Hz    & Repetition rate & 400 Hz \\
 
 \hline
\end{tabular}
    \caption{Simulation parameters used to model the production of 975-keV $\gamma$ rays with an on-axis RMS energy bandwidth of 1\% and a total flux of $2.3 \times 10^{12}$ photons/second through laser-Compton scattering of directly counter-propagating beams. These parameters are based on DCCS design specifications for the DARPA Gamma Ray Inspection Technology (GRIT) program. Both the laser and electron beam are modeled to be radially symmetric about their propagation axes. $\sigma_i$ refers to the standard (RMS) deviation of the underlying distribution. $\varepsilon_{n, rms}$ is the transverse normalized RMS emittance of the electron beam. FTL = Fourier transform limit.}
    \label{table:modeling}
\end{table}

The X-band electron accelerator system for the DCCS under development at Lumitron Technologies, Inc. is commissioned to produce low-emittance electron beams with energies up to 137 MeV as part of the DARPA Gamma Ray Inspection Technology (GRIT) program.
Based on previous work at Lawrence Livermore National Laboratory (LLNL), the SLAC National Accelerator Laboratory, and further developments to the X-band accelerator systems at Lumitron Technologies (details in Sections \ref{sec:photogun} and \ref{sec:linacs}), normalized electron beam transverse emittance values of 0.3~mm-mrad are expected.
The beamline at Lumitron Technologies, Inc. is designed to focus such an electron beam to an RMS-radius focal spot of 1.5~\textmu m with an RMS bunch length of 0.59~ps.

In the numerical analysis presented here, we consider a UV interaction laser with a central wavelength of 345.67~nm and a FWHM pulse length of 2.0~ps produced through third-harmonic generation from a 1064-nm Nd:YAG laser.
Though the experimental results presented later in this work use a 532-nm interaction laser (Sections \ref{sec:compton-results}), the simulation results presented here are still valid with the consideration that the final laser-Compton scattered energy is linearly proportional with the frequency of the interaction photons.
With these values set, we leave the transverse focal size of the interaction laser as a free parameter to solve for the size that maximizes total Compton-scattered photon flux.

\begin{figure}[ht!]
    \centering
    \includegraphics[width=17.5cm]{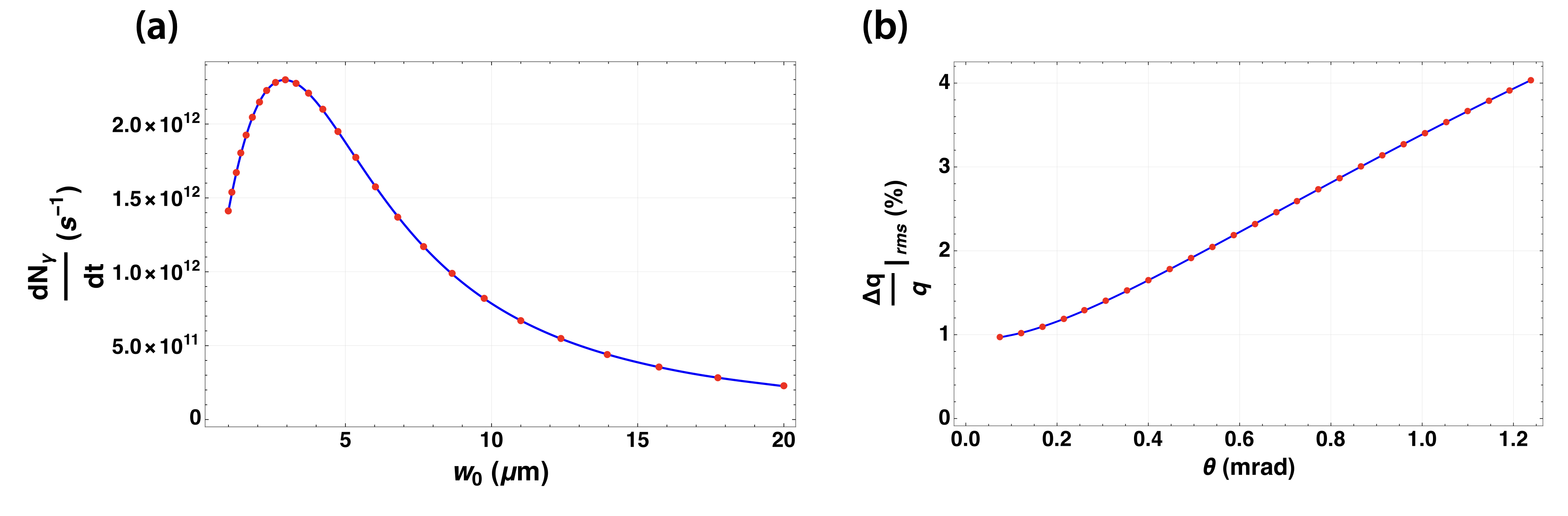}
    \caption{(a) Calculation of Compton-scattered photons produced per second from counter-propagating electron and laser pulse trains following the parameters summarized in Table \ref{table:modeling} as a function of the $1/e^2$-radius of the focused interaction laser ($w_0$). 2.94~\textmu m (3.46~\textmu m FWHM) is identified as maximizing the total flux of the interaction at $2.3 \times 10^{12}$ photons per second. (b) Corresponding relative bandwidth of the resulting Compton-scattered $\gamma$-ray beam as a function of integrated solid angle about the central beam axis.}
    \label{fig:modeling}
\end{figure}

The hallmark of the DCCS architecture is the distribution of electron bunch charge and laser pulse energy over long pulse trains (macrobunches) that are produced at high repetition rates.
The commissioning goals of the DCCS at Lumitron are to produce electron macrobunches and laser macropulses with bunch-train lengths of 1000 bunches and 100 pulses respectively.
This mismatch will ultimately be overcome by using a laser recirculation strategy based on previously described work first pioneered at LLNL \cite{shverdin:2010:OL}, in which a 10x enhancement in effective laser/electron overlap is expected.
With this recirculation cavity in place and with both the electron beam and interaction laser systems operating at 400 Hz, an expected 400,000 interactions are expected to occur each second. This scaling factor of $4 \times 10^5$ is included in the flux calculations shown in Figure \ref{fig:modeling}.

Using a previously described numerical model \cite{albert:2011:PoP}, a three-dimensional diffracting Gaussian laser pulse is overlapped with a relativistic electron beam considering its full six-dimensional phase space.
This model also accounts for electron recoil and potential nonlinear effects induced by the ponderomotive force during interaction while calculating the resulting spectrum of Compton-scattered photons.
The simulation inputs are based on the parameters summarized in Table \ref{table:modeling}.
Assuming a counter-propagating laser-electron geometry, a 3.46-\textmu m FWHM diameter was found to be the optimum focal spot size for the interaction laser, with a total expected output flux of $2.3 \times 10^{12}$ photons per second (Figure \ref{fig:modeling}(a)), a peak on-axis energy of 975 keV, and an on-axis RMS energy bandwidth of 1\% (Figure \ref{fig:modeling}(b)).

This numerical model provides the theoretical foundation of the DCCS architecture and its suitability for producing both narrow bandwidth x rays (or $\gamma$ rays) and VHEE beams.
In the systems integration results reported in Section \ref{sec:compton-results}, the working interaction laser and electron beam specifications are summarized in Table \ref{table:stats}.
The operational conditions of the electron beam during the results reported in Section \ref{sec:vhee-results} are also summarized in Table \ref{table:stats}.
Both electron and interaction laser systems were operating at 100~Hz, with a typical electron bunch charge of 5~pC and an interaction laser pulse energy of 2.5~mJ and electron energies around 40 MeV.
Typical normalized emittance values measured were below 0.5~mm-mrad, with electron bunch lengths indirectly measured to be 1.1~ps RMS based on streak camera measurements of the photoinjector laser after conversion to UV.
Interaction laser pulse lengths were measured to be 7~ps FWHM.
Additionally, since no recirculation cavity was installed at the time of the presented experiments, the electron bunch train lengths were set to 100 to match those of the interaction laser.
In the results presented in Section \ref{sec:vhee-results}, the electron beam was set to produce 1000-bunch trains at 100~Hz. Finally, while the interaction laser was focused to a 5-\textmu m FWHM spot, the electron beam was only focused to a 17-\textmu m RMS spot (41-\textmu m FWHM) to facilitate alignment during this first experimental campaign.

\subsection{High-current, high-brightness photoguns}
\label{sec:photogun}

The minimum on-axis bandwidth from a laser-Compton system depends strongly on emittance of the accelerated electron bunches \cite{Sun:2009:PRSTAB, krafft:2016:PRAB, hwang:2023:PRAB}.
The production of low-emittance electron beams ($\epsilon_n = 0.35$~mm-mrad at 20 pC/bunch and $\epsilon_n = 0.8$ mm-mrad at 100 pC/bunch) has previously been demonstrated using a 5.5 cell X-band photogun (Mark 0) \cite{limborg-deprey:2016:PRAB}, based on a design by LLNL and SLAC \cite{marsh:2012:PRSTAB}.
Later, the Mark~1 version of this X-band photogun (5.59~cells) demonstrated significantly improved beam emittance relative to Mark~0 at substantial charge per bunch ($\epsilon_n = 0.3$~mm-mrad at 80 pC/bunch) \cite{marsh:2018:PRAB}.
This LLNL/SLAC X-band photogun concept was originally designed to operate in a single bunch mode, with a nominally 250-pC bunch charge.
At LLNL, the feasibility of using the LLNL/SLAC X-band photogun with distributed charge operation was supported by initial modeling studies \cite{marsh:2012:PRSTAB, marsh:2012:IPAC}, and was demonstrated experimentally with 11.424-GHz bunch trains using up to 4 \cite{gibson:2016:EUVXRAY} and then up to 16 consecutive electron bunches \cite{gibson:2017:IPAC}.
These initial demonstrations of 11.424 GHz distributed charge operation did not use pulse synthesis, but rather, a hyper-Michelson interferometer to space the photogun drive laser pulses by 87.5~ps \cite{siders:1998:AO}.
Additionally, in the first demonstrations of multi-bunch operation, the photogun drive laser operated at 10 Hz \cite{gibson:2016:EUVXRAY, gibson:2017:IPAC}.
The integration of pulse synthesis (Section \ref{sec:pulse-synthesis}) in the DCCS architecture potentially extends these results to 1000 consecutive electron bunches with repetition rates up to 400 Hz.

\begin{figure}[ht!]
    \centering
    \includegraphics[width=17cm]{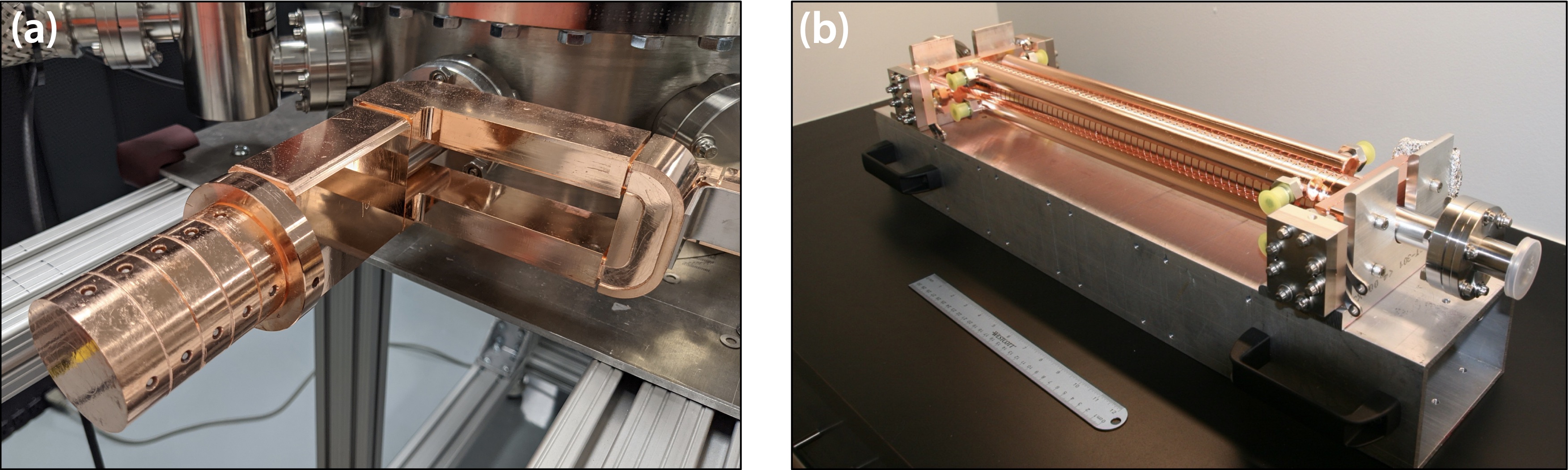}
    \caption{(a) X-band photogun designed, fabricated, and tuned at Lumitron Technologies, Inc. (b) T53VG3 X-band accelerator section fabricated and tuned by Lumitron Technologies, Inc.}
    \label{fig:t53}
\end{figure}

This extension to long pulse trains and high repetition rates presents two fundamental challenges.
The first challenge is the uniform acceleration of all 1000 bunches during one RF pulse, and the second challenge is controlling the thermal loading and distortions that accompany the higher repetition rate.
The former requires precision shaping of the RF pulses that drive the gun and accelerator sections and is further complicated by the use of RF pulse compression on the output of each klystron/modulator unit.
Low power tests to date using shaped RF seed pulses provide encouragement that the required RF pulse shapes and stability can be achieved at high power.
Thermal models for operation of the gun at 400 Hz indicate that mechanical distortions of the gun cells would be significant enough to impact performance if not compensated.
To alleviate this issue, a custom cooling jacket has been designed and constructed for the next generation of X-band photoguns to handle high repetition rate operation.
These novel X-band photoguns rated for high repetition rate operation are fabricated from oxygen-free copper.
Parts were machined to 1-\textmu m accuracy, diffusion bonded, brazed, RF tuned, and prepared for ultra-high vacuum (UHV) installation at Lumitron Technologies (Figure \ref{fig:t53}(a)).

\subsection{High-gradient X-band LINACs}
\label{sec:linacs}

In the DCCS, high-gradient, X-band (11.424 GHz) LINACs are operated in a distributed charge mode with average beam currents of up to 10 \textmu A.
The baseline specification is 25 pC per microbunch with 1000 microbunches per macrobunch and a macrobunch repetition rate of 100 Hz (2.5 \textmu A) with potential operation up to 3000~microbunches and 400 Hz (30 \textmu A).
This baseline configuration represents an extension of X-band multi-bunch operation from 16 microbunches per macrobunch demonstrated at LLNL \cite{gibson:2017:IPAC} up to 3000.
In this work, we demonstrate successful multi-bunch operation with electron beam quality high enough to produce Compton-scattered x rays using 100-bunch trains at 100 Hz (Section \ref{sec:compton-results}).
The T53VG3 accelerator design was chosen because of its technological maturity, its demonstration of exceptional acceleration gradients greater than 100 MeV/m \cite{adolphsen:2003}, and its use of a traveling wave accelerating field, making it less susceptible to electron beam-induced electromagnetic wake fields.

Regarding the accelerator sections, the primary concerns with this approach are variations in microbunch energy and emittance within the bunch train and overall bunch train pointing stability due to electron-induced wakes.
Modeling to date suggests that the electron bunch wakes should be minimal from 1000 bunch, multi-bunch operation at 25 pC per microbunch.
Further modeling has also informed the exact RF pulse shapes that will be required to drive both the photogun and the LINAC sections so that all bunches emerge from the system with the same energy ± 0.1\%.
The chosen X-band photogun and high-gradient accelerator technologies are higher beam current extensions of the X-band (11.424 GHz) designs previously demonstrated both at the Lawrence Livermore National Laboratory and the SLAC National Accelerator Laboratory.
These higher-current systems have increased thermal loading and require the addition of precision thermal management.
The T53VG3 LINAC sections in this work were fabricated from oxygen-free copper. Parts were machined to 2.5-\textmu m accuracy, RF tuned, and prepared for UHV installation at Lumitron Technologies (Figure \ref{fig:t53}(b)).

\subsection{RF laser-pulse synthesis and amplification of 11.424 GHz pulse trains}
\label{sec:pulse-synthesis}

\begin{figure}[ht!]
    \centering
    \includegraphics[width=17.5cm]{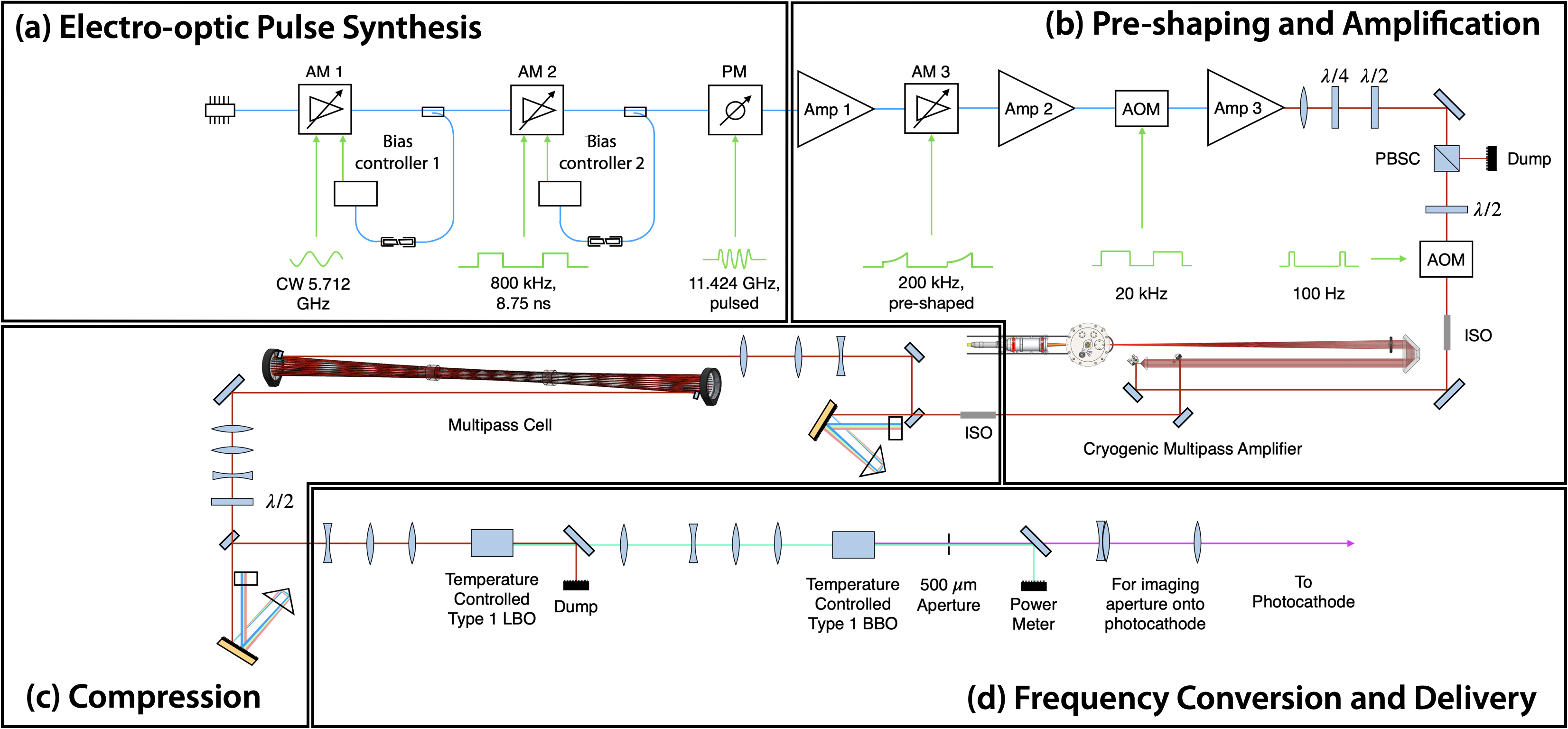}
    \caption{Diagram of the photogun laser system (PGL) used at Lumitron Technologies, Inc. for the results presented in this work. (a) Pulse synthesis approach used to generate 11.424 GHz micropulse trains from a 995.2~nm CW seed laser. A CW laser is first amplitude modulated (AM 1) with a 5.712 GHz sinusoidal signal. The resulting 11.424 GHz optical signal is then partitioned into macropulses with an 800-kHz signal from a bit pattern generator (AM 2). Finally, the micropulse bandwidth is broadened with a comb line structure after sinusoidal phase modulation (PM). (b) The micropulse trains are amplified (Amp 1), preshaped and reduced in repetition rate to 200~kHz (AM 3), amplified again (Amp 2), reduced to 20~kHz using an AOM, then sent through a final fiber amplifier (Amp 3) before being sent through a final AOM to reduce repetition rate to 100~Hz. The pulse trains are then sent through a multi-pass bulk amplifier system. (c) The fully amplified pulse trains are sent through a grating compressor, spectrally broadened through a multi-pass cell, and compressed again through a second grating compressor. (d) The pulse trains are frequency converted to the \nth{4} harmonic (249~nm) before being imaged from a 500~\textmu m aperture onto the photocathode for electron production. CW = continuous wave, AM = amplitude modulator, PM = phase modulator, Amp = amplifier, AOM = acousto-optic modulator, PBSC = polarizing beam splitter cube, ISO = optical isolator, LBO = lithium triborate, BBO = barium borate.}
    \label{fig:pulse-synthesis}
\end{figure}

An inherent challenge for any laser-Compton source is the synchronization of the arrival of the electrons and laser photons at a common focus.
In all existing systems, the accelerator RF (3 to 12 GHz) is significantly greater than the natural repetition rate of short-duration, mode-locked lasers (80 to 100 MHz) that seed the photogun and interaction laser systems.
Even specialized systems used at CLEAR that use a 1.4-GHz mode-locked laser still need to utilize an interferometer scheme to maximally distribute charge over a train of consecutive bunches at 3 GHz \cite{granados:2019}.
Timing synchronization is usually accomplished via locking of the laser repetition rate to a sub-multiple of the accelerator RF via precision movement of an intra-cavity, piezoelectric-actuated element.
The whole system is then controlled by one or more phase locked loops.
The DCCS architecture takes a fundamentally different approach to the synchronization problem by synthesizing the drive laser pulses using the accelerator RF.
In this patented approach \cite{messerly-patent} that has been previously demonstrated at 11~GHz \cite{prantil:2013:OL}, a stable, narrow-bandwidth seed laser is phase and amplitude modulated using standard, 40-GHz bandwidth, telecommunications-quality fiber optic components to produce trains of approximately 50-ps duration, chirped, IR laser pulses that then seed both the photogun laser system (PGL) and the interaction laser system (ILS).
This section will, as an example, focus on the pulse synthesis (Figure \ref{fig:pulse-synthesis}(a)) and pre-amplification (Figure \ref{fig:pulse-synthesis}(b)) stages of the PGL.

The output of a continuous wave (CW) laser diode is first amplitude modulated using a 5.712~GHz signal with a null bias, resulting in an 11.424~GHz laser output.
This initial amplitude modulation defines the structure of the laser micropulses, and is performed using the same RF input that is used as a seed for the RF power systems that drive the accelerator system.
An 800~kHz signal then amplitude modulates the signal to carve the macropulse structure.
For the laser-Compton x-ray results presented here, the width of this signal was 8.75~ns, corresponding to 100~micropulses per macropulse.
For the electron beam results presented here, the width of this signal was adjusted to 87.5~ns, corresponding to 1000~micropulses per macropulse.
Finally, an 11.424~GHz RF signal is used to phase modulate, and thus spectrally broaden, the laser micropulse trains.
Critically, the RF input to the phase and amplitude modulators is the same RF that is used as a seed for the RF power systems that drive the accelerator system.
In this way it is possible to create a train of laser pulses whose repetition rate matches exactly to the bunch repetition rate of the accelerator system.
Any phase instability in the seed RF is transferred automatically to both systems identically.
Thus, the Compton interaction timing problem is reduced to establishing a simple optical or electronic delay for the laser pulses illuminating the photocathode and used in Compton x-ray generation.

\begin{figure}[ht!]
    \centering
    \includegraphics[width=17.5cm]{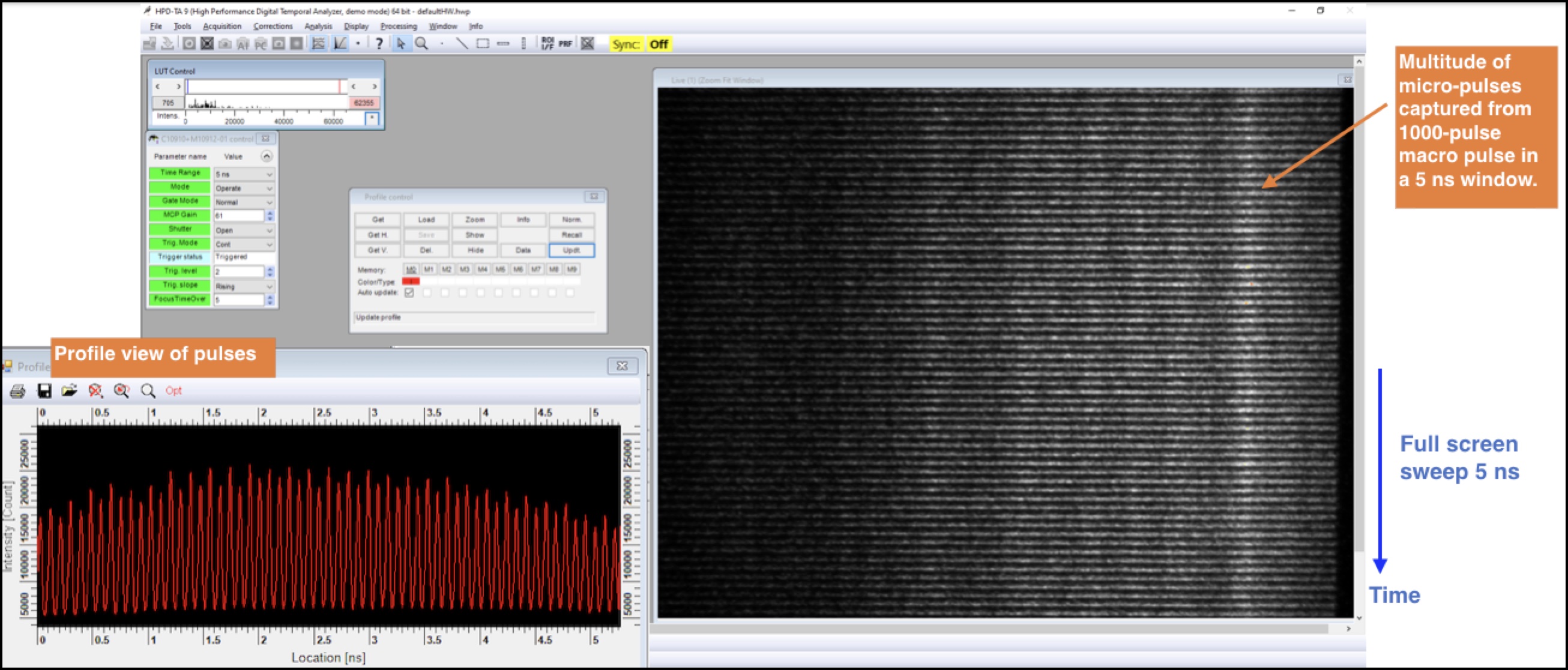}
    \caption{Streak camera recording of evenly-spaced micropulses within a single laser macropulse used to produce electron bunches at the DCCS photocathode. 59 micropulses are present within the illustrated streak camera exposure window of 5.2~ns. This is consistent with a micropulse spacing of 87.5~ps which is in turn set by the master clock operating frequency of 11.424~GHz. Illustrated at the left is an integration over the horizontal dimension of the raw streak camera image.}
    \label{fig:streak-camera}
\end{figure}

Experimental verification of the production of 11.424~GHz micropulse trains is presented in Figure \ref{fig:streak-camera}.
The right portion of this figure shows a raw streak camera trace of 59~micropulses taken from the PGL system captured within a 5.2~ns window.
On the left side of the figure is a profile view of the streak trace integrated over the horizontal dimension.
The 5.2~ns window was chosen to maximize the number of micropulses that could be seen while maintaining enough resolution to visualize the individual macropulses.
For the experimental results presented in Section \ref{sec:compton-results}, micropulse durations were measured to be 2.5~ps FWHM.

The initial amplification of the PGL and ILS also occur in fiber, going through three amplification stages as the macropulse repetition rate is eventually decreased to 20~kHz (PGL) or 200~kHz (ILS).
Before entering the bulk amplification stage, which will be discussed in Section \ref{sec:infrared}, the repetition rate is reduced to its final operational value using acousto-optic modulators.
For the results presented here, this final operational value was 100~Hz.

\subsection{Diode-pumped infrared laser technology}
\label{sec:infrared}

\begin{figure}[ht!]
    \centering
    \includegraphics[width=17.5cm]{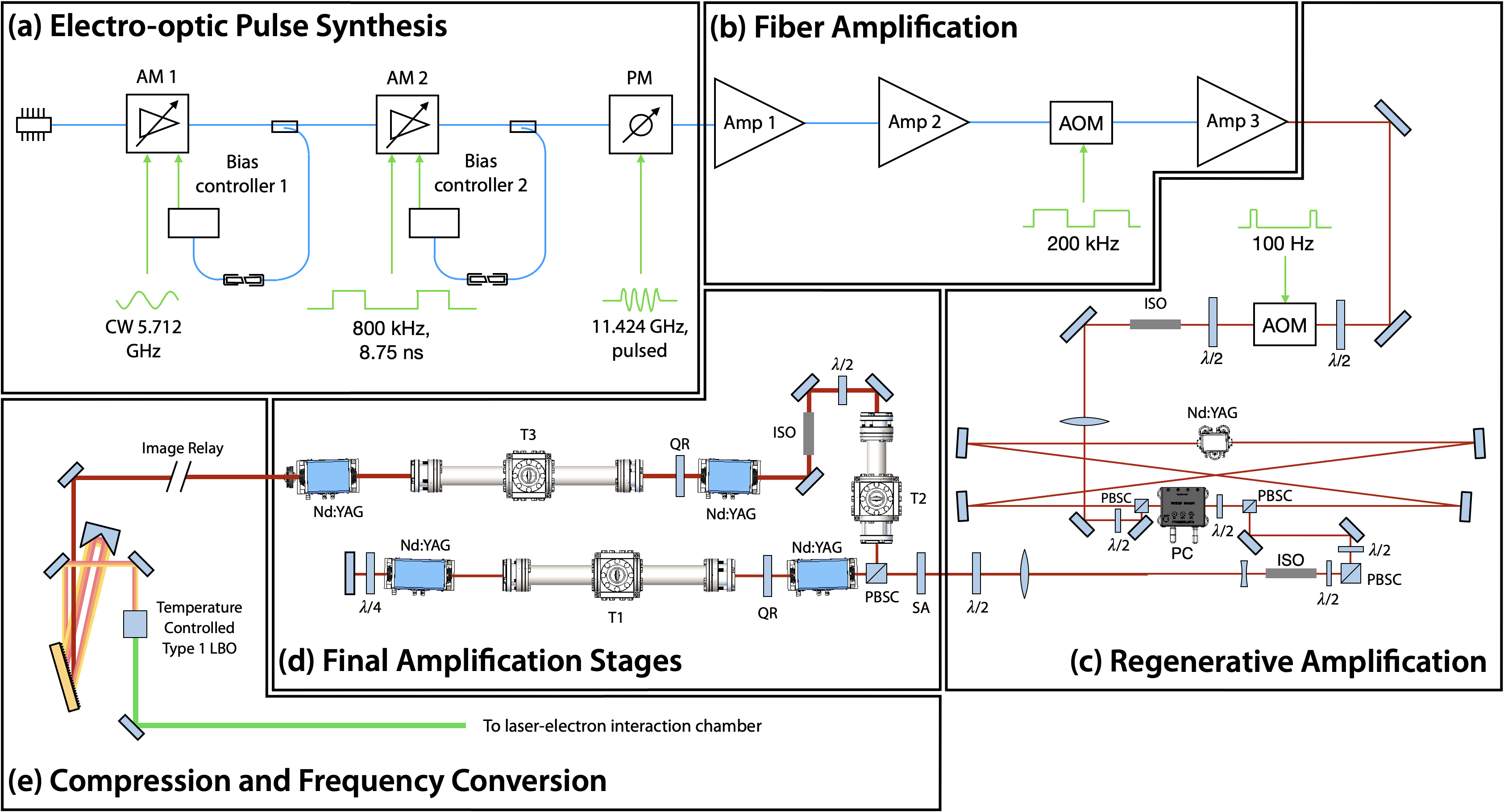}
    \caption{Diagram of the interaction laser system (ILS) used at Lumitron Technologies, Inc. for the results presented in this work. (a) Pulse synthesis approach used to generate 11.424 GHz micropulse trains from a 1064~nm CW seed laser. A CW laser is first amplitude modulated (AM 1) with a 5.712 GHz sinusoidal signal. The resulting 11.424 GHz optical signal is then partitioned into macropulses with an 800-kHz signal from a bit pattern generator (AM 2). Finally, the micropulse bandwidth is broadened with a comb line structure after sinusoidal phase modulation (PM). (b) The micropulse trains are amplified twice (Amp 1 and Amp 2) before being reduced in repetition rate to 200~kHz (AM 3), and amplified again (Amp 3). (c) The free-space macropulse repetition rate is reduced to 100~Hz with an AOM before being sent into a Nd:YAG regenerative amplifier. The PC switches the polarization of the pulse train after about 30 passes. (d) The amplified and stretched pulse train is then double-passed through two 7-mm diameter Nd:YAG crystals and single-passed through two 12-mm diameter Nd:YAG crystals as the final amplification stages. (e) The pulse trains are temporally compressed with partial spatial overlap, frequency converted to the \nth{2} harmonic (532~nm), and finally delivered to the laser-electron interaction chamber. CW = continuous wave, AM = amplitude modulator, PM = phase modulator, Amp = amplifier, AOM = acousto-optic modulator, PBSC = polarizing beam splitter cube, PC = Pockels cell, Nd:YAG = neodymium-doped yttrium aluminum garnet, ISO = optical isolator, QR = 90$^\circ$ quartz rotator, SA = serrated aperture, T1-3 = vacuum telescopes with spatial filters, LBO = lithium triborate.}
    \label{fig:ils-system}
\end{figure}

Two laser amplification systems are used to produce the results discussed in this work (Sections \ref{sec:compton-results} and \ref{sec:vhee-results}), both of which are based on diode-pumped IR laser technologies.
The laser amplifier system used to produce the ILS that is ultimately focused and collided with a counter-propagating electron beam is a Nd:YAG regenerative amplifier and multipass bulk amplifier system that amplifies a highly structured seed pulse (see Section \ref{sec:pulse-synthesis}) at 1064~nm.
A schematic of the ILS is shown in Figure \ref{fig:ils-system}.
In the systems integration results for the production of laser-Compton x rays reported here, the ILS is frequency converted to produced 532~nm macropulses at 100~Hz (Table \ref{table:stats}).
Each macropulse consisted of 100 micropulses spaced at 87.5~ps, corresponding to the 11.424~GHz LINAC operating frequency.
These highly structured pulses were generated through electro-optic pulse synthesis (Figure \ref{fig:ils-system}(a)), fiber amplification (Figure \ref{fig:ils-system}(b)), and subsequent repetition rate reduction to 100~Hz after passing through an acousto-optic modulator (AOM) in free space.
The ILS micropulses were then amplified through a diode-pumped Nd:YAG (4-mm diameter) regenerative amplifier (Figure \ref{fig:ils-system}(c)).
After 30 passes, the macropulse is dumped from the regenerative amplifier cavity using a Pockels cell.
In the final stages of amplification, the ILS pulses were double-passed through two 7-mm diameter diode-pumped Nd:YAG rods and single-passed through two 12-mm diode-pumped Nd:YAG rods (Figure \ref{fig:ils-system}(d)).
During the final amplification stages, the pulses were sent through three Keplerian vacuum  telescopes with spatial filters labeled T1, T2, and T3 in Figure \ref{fig:ils-system}(d).
After the ILS pulses were fully amplified, they were image relayed to a grating compressor where the pulses were temporally compressed before \nth{2}~harmonic frequency conversion.
After \nth{2} harmonic conversion to 532~nm, 25~W of average power was available for the laser-Compton interaction.
Considering 100 micropulses per macropulse at a 100-Hz macropulse repetition rate, this corresponded to an average micropulse energy of 2.5~mJ.

Assuming a copper photocathode quantum efficiency of $10^{-5}$ and 1000 microbunches, an average IR power of nominally 10~W would be needed for VHEE FLASH operation.
Thus, Yb:YLF was chosen as an initial photogun laser (PGL) amplifier system (Figure \ref{fig:pulse-synthesis}(b)) using a previously reported cryogenic Yb:YAG multi-pass amplifier design \cite{zapata:2022}.
The multi-pass amplifier system from Zapata, \textit{et al.} (2023) \cite{zapata:2022} was modified to operate with Yb:YLF at the 995.2~nm emission line \cite{Zapata:2010:995nm}.
Operation at 995.2~nm decreases the quantum defect when pumping with high-power 960~nm diodes, and enables the potential to operate at greater than 10~W of average power by decreasing cooling requirements on the crystal.
One limitation of this PGL design is that the 995.2~nm emission line is narrower than the emission band near 1020~nm.
To overcome gain narrowing and reach sufficiently short pulses to maintain electron beam quality, a multi-pass nonlinear compression scheme was implemented to spectrally broaden the pulse (Figure \ref{fig:pulse-synthesis}(c)).
This allowed for subsequent grating compression to pulse widths as low as 2.5~ps before \nth{4} harmonic frequency conversion and incidence onto the photocathode surface (Figure \ref{fig:pulse-synthesis}(d)).
This nonlinear compression scheme employed during production of laser-Compton x rays (see Section \ref{sec:compton-results}) was an in-air multi-pass cell optimized for low pulse energies and capable of compressing pulses with initial pulse widths as long as 14~ps \cite{seggebruch:2024}.
To maximize charge production at the photocathode, the multi-pass cell was bypassed, resulting in degraded electron beam quality during the electron beam studies in Section \ref{sec:vhee-results}.

It should be noted that Lumitron’s patented DCCS architecture distributes the energy of the laser macropulse over 100 micropulses and as such reduces the peak intensity of laser pulses transmitted through optical windows, lenses, and nonlinear crystals by 2 orders of magnitude relative to laser-Compton architectures based on a single high-energy, short duration interaction laser pulse.
This eliminates the potential for optical damage from nonlinear self-focusing and significantly expands the design possibilities for the optics within the overall system.
For example, this architecture enables the use of lenses (as opposed to curved mirrors) for beam focusing and beam transport which would otherwise not be possible for higher peak intensity pulses.

\begin{table}[ht!]
    \centering
\begin{tabular}{ p{4cm}p{4cm}|p{4cm}p{4cm} }
 \hline
 \multicolumn{2}{c|}{\textbf{Interaction Laser}} &
 \multicolumn{2}{c}{\textbf{Electron Beam}}\\
 \hline
 Wavelength         & 532 nm & Beam energy & 38 (49.4) MeV \\
 Micropulse energy  & 2.5 mJ     & Bunch charge &   5 (14) pC \\
  $M^2$ & $\sim 1.6$ &  $\varepsilon_{n, rms}$* &   $<$ 0.5 mm-mrad \\
 FTL bandwidth\textsuperscript{\textdagger}  & 60~pm   &Energy spread* ($\sigma_E / E$)  &   0.2\% \\
 Pulse train length & 100     & Bunch train length & 100 (1000)\\
 Recirculation efficiency  & N/A    & Duration ($\sigma_t$) & 1.1 ps \\
 FWHM duration   & 7 ps    & Duration ($\sigma_\theta)$ &  4.52$^{\circ}$ \\
 FWHM diameter     & 5 \textmu m  & Radius ($\sigma_x$, $\sigma_y$) &  17 \textmu m, 19 \textmu m \\
 Pulse spacing      & 87.535 ps  & X-band frequency & 11.424 (11.424) GHz \\
Repetition rate    & 100 Hz & Repetition rate & 100 (100) Hz \\
 
 \hline
\end{tabular}
    \caption{Measured interaction laser and electron beam parameters used to produce the experimental results presented in this work. For the electron beam parameters, numbers in parentheses indicate values that were measured for Section \ref{sec:vhee-results}. Parameters without corresponding parenthetical values were not measured during those Section \ref{sec:vhee-results} experiments. *Here, these values refer to the effective values when considering the entire 100-bunch train (macrobunch) of electrons. \textsuperscript{\textdagger}The interaction laser FWHM bandwidth was only directly measured in the IR (260 pm) before frequency conversion. The listed value is the FTL FWHM bandwidth assuming an underlying Gaussian spectrum. It can serve as a lower bound on the bandwidth based on the measured 532-nm FWHM pulse duration. FTL = Fourier transform limit.}
    \label{table:stats}
\end{table}

\section{Laser-Compton X-ray Results}
\label{sec:compton-results}

\begin{figure}[ht!]
    \centering
    \includegraphics[width=17.5cm]{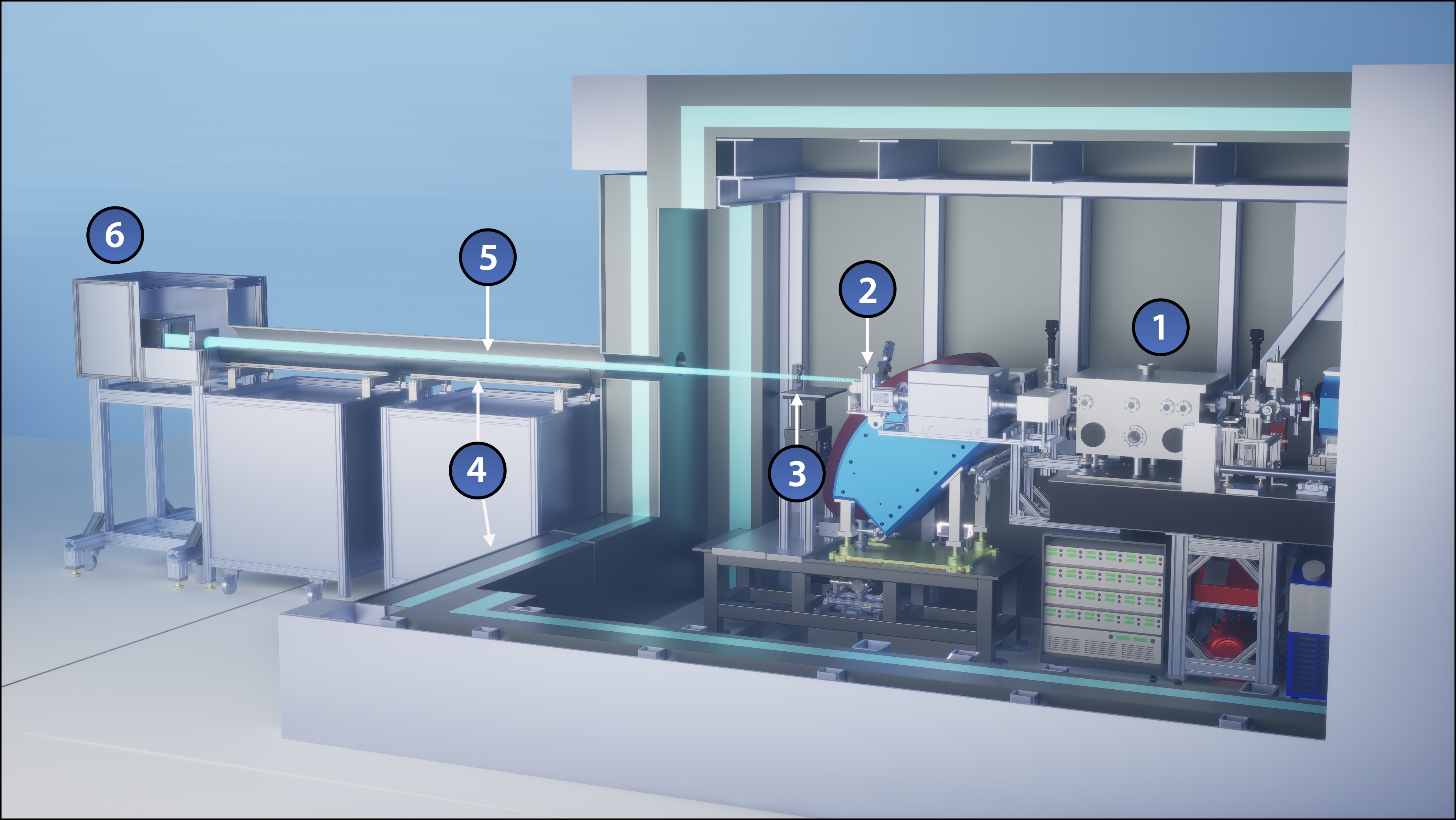}
    \caption{Laser-Compton x-ray imaging setup at Lumitron Technologies, Inc. (1) Laser-Compton interaction chamber where laser pulses scatter off of counter-propagating electron bunches to produce a pulsed, laser-Compton x-ray beam. (2) X-ray output window where x rays exit vacuum and begin propagating through air. (3) Object to be imaged is placed on a translation stage system to align with the x-ray beam. (4) Radiation shielding is used to house the accelerator and beam dump in the bunker enclosure in addition to a Pb-lined beam tube that surrounds the x-ray beam as it leaves the bunker. (5) The Laser-Compton x-ray beam. (6) A high resolution detector system is placed in a shielded hutch to detect LCS x rays.}
    \label{fig:imaging-system}
\end{figure}

As part of the DCCS systems integration test reported here, the laser-Compton x-ray beam produced using the DCCS architecture was tested for feasibility of precision imaging.
Figure \ref{fig:imaging-system} provides a detailed look at the imaging setup used in the systems integration test.
After laser-Compton x rays are produced in the laser-electron interaction chamber, the x rays leave vacuum and pass through an experimental bay dedicated to holding sample objects for imaging.
After passing through objects of interest, the imaging x rays propagate out of the radiation safety bunker, through a modular set of beam tubes, and finally onto an x-ray imaging detector.
For the results presented in this work, a flat panel x-ray detector was used (Varex, 1512 CMOS), with a pixel pitch of 74.8~\textmu m, a field-of-view of 15~cm by 12~cm, and 200~\textmu m of CsI as the scintillation material.

Laser-Compton x-ray beam profiles were imaged with peak on-axis energies ranging from 30~keV to 140~keV.
During these experiments, the x-ray source volume was predominantly determined by the ILS spot size when focused at the interaction point.
While the beam quality ($M^2$) was not measured directly, the measured focal spot is consistent with an equivalent-diameter 8th-order super-Gaussian assuming a flat spatial phase at the input to the lens.
To facilitate alignment for these first imaging demonstrations, the electron beam was focused to a 17-\textmu m RMS spot (41-\textmu m FWHM) at the interaction point.
The interaction laser was operated at 25~W of average power while producing trains of 100~micropulses at 100~Hz. This corresponds to 2.5~mJ per micropulse.

\begin{figure}[ht!]
    \centering
    \includegraphics[width=15cm]{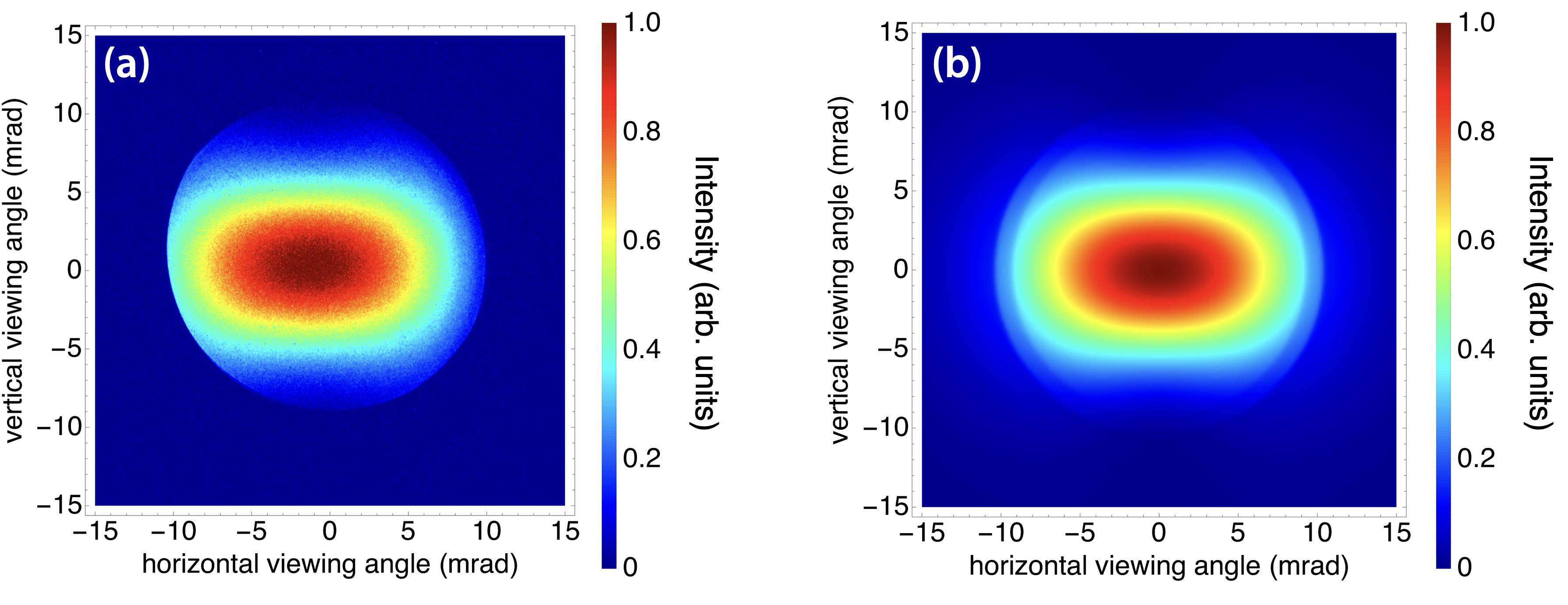}
    \caption{(a) An x-ray image produced using $10^{10}$ photons and (b) the corresponding simulation  for laser-Compton x-ray beam tuned to a peak on-axis energy of 54 keV. The shape of the x-ray intensity distribution is consistent with a dipole radiation pattern resulting from a vertically polarized interaction laser. The apparent cutoff in the distribution pattern in (a) is due to beam offset relative to the vacuum window aperture.}
    \label{fig:compton-beam}
\end{figure}

\begin{figure}[ht!]
    \centering
    \includegraphics[width=17.5cm]{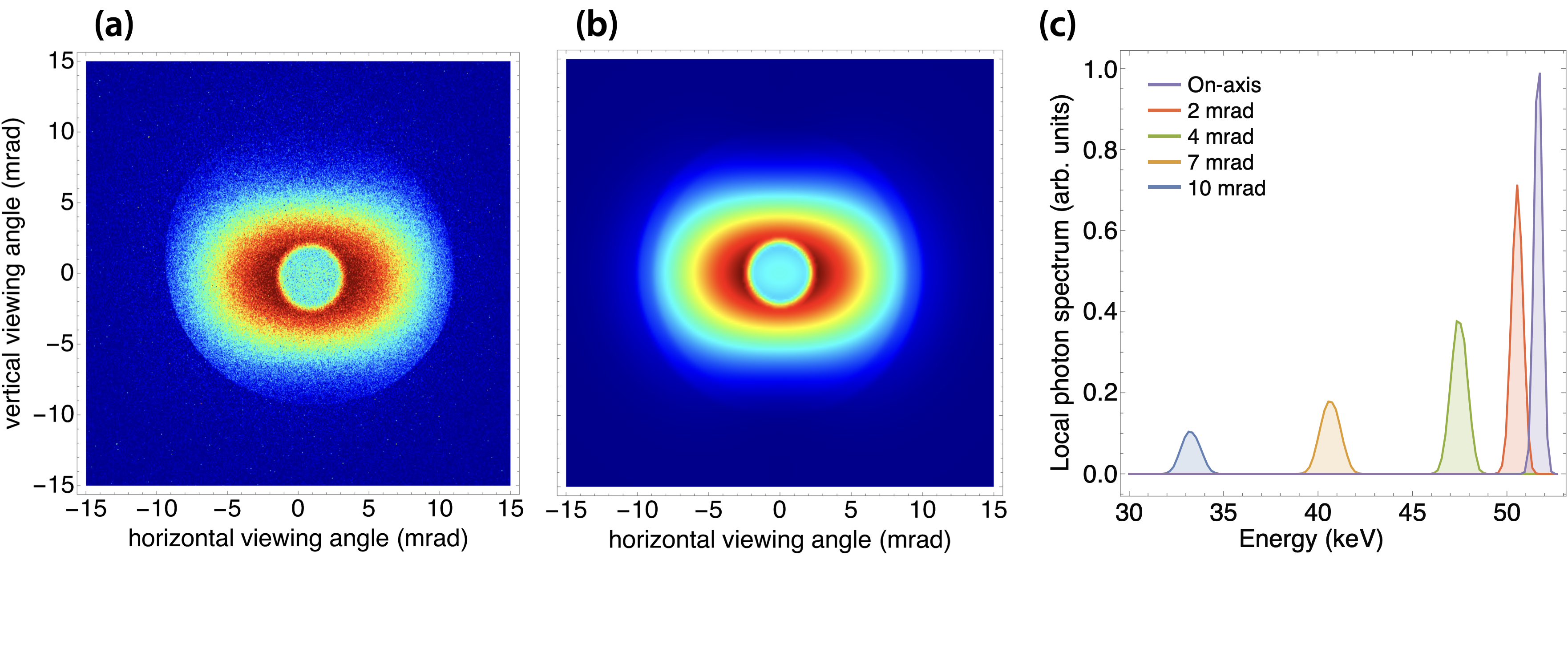}
    \caption{(a) An x-ray image and (b) corresponding simulation for laser-Compton x-ray beam tuned to a peak on-axis energy of 51.8 keV. The beam is sent through a 100-\textmu m-thick Gd foil before reaching the detector, demonstrating the angle-correlated energy spectrum of the laser-Compton x rays. (c) Numerically reconstructed local energy spectra of the best-fit simulation, which used the measured electron beam energy spread of 0.2\% to predict an on-axis x-ray RMS bandwidth of 0.41\%.}
    \label{fig:gd-k-hole}
\end{figure}

The interaction of a 532~nm (2.33~eV) laser pulse train with an 38.5~MeV electron beam was used to produce laser-Compton scattered x-ray photons with a peak on-axis energy of 54~keV shown in Figure\ref{fig:compton-beam}(a).
The image shown is the accumulation of $10^{10}$ x rays.
Figure \ref{fig:compton-beam}(b) is a simulated x-ray beam using the same expected interaction parameters.
In order to demonstrate the angle-correlated spectrum and determine the on-axis bandwidth of the laser-Compton x rays, a 100-\textmu m-thick Gd foil was placed in the beamline and the peak on-axis was tuned to 51.8~keV by changing the electron energy to 37.4~MeV (Figure \ref{fig:gd-k-hole}(a)).
x rays with energies just above the K-edge of Gd will be highly attenuated compared to energies just below the K-edge.
Since the K-shell absorption edge of Gd is 50.2 keV, and since the spectrum of a laser-Compton source produced using a low-emittance electron beam is highly angle-correlated, an attenuation ``hole'' will appear where the mean energy of the angle-correlated energy spectrum is above the K-edge near the center of the beam.
Previous work has demonstrated that the sharpness of this hole is related to the divergence and energy spread of the electron beam \cite{hwang:2023:PRAB}.
The electron beam parameters that corresponded with the simulated x-ray beam profile that best matched the experimental x-ray spectrum were consistent with the measured electron beam energy spread of 0.2\% and expected divergence of around 0.12~mrad (Figure \ref{fig:gd-k-hole}(b)).
Based on this approach, the best-fit simulated x-ray spectrum corresponds to an on-axis RMS energy bandwidth of 0.41\%.
This is consistent with a relatively large electron beam that is not focused as sharply as that used to produce the simulated results in Figure \ref{fig:modeling}.
To our knowledge, this is the narrowest on-axis bandwidth ever produced from a normal-conducting laser-Compton source.

\begin{figure}[ht!]
    \centering
    \includegraphics[width=17.5cm]{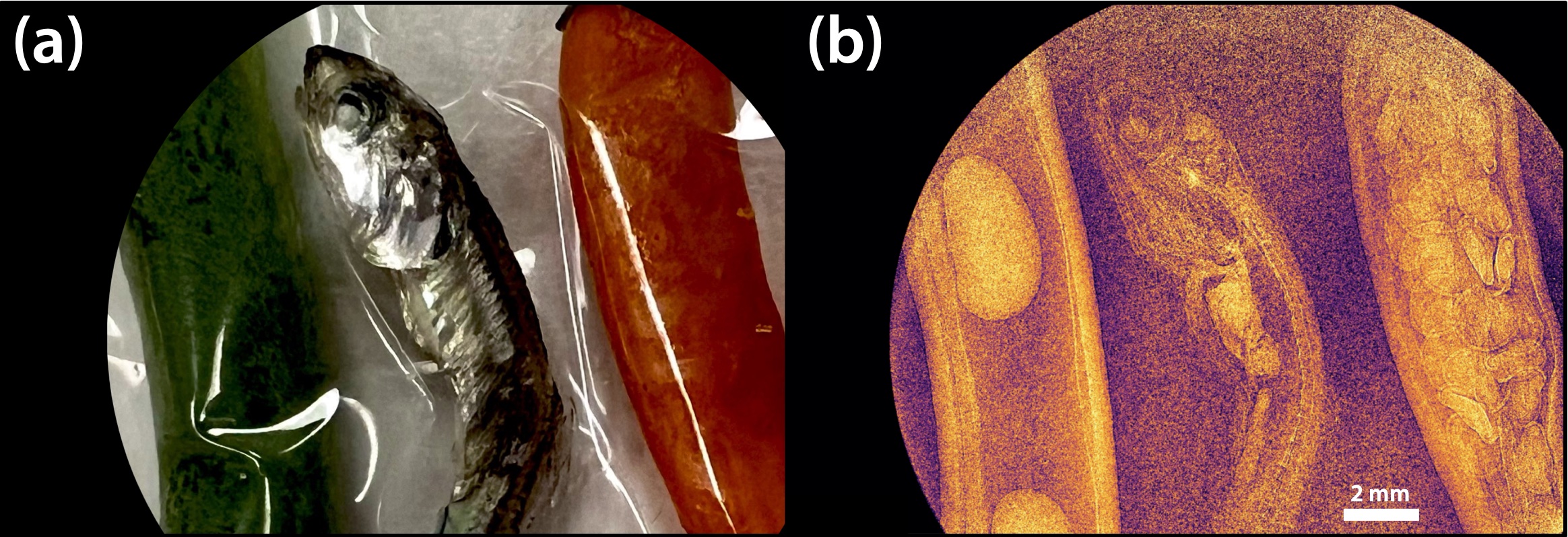}
    \caption{(a) Photograph of, from left to right, a bean pod, a dried anchovy, and a chili pepper. The image is vignetted to match the field of view of the corresponding x-ray image. (b) Laser-Compton x-ray image of the photographed objects using a total of $3.2 \times 10^{10}$ x-ray photons with a peak on-axis energy of 50~keV. Phase-based edge enhancement is visible on various structures, including the seeds inside the chili pepper. The width of the anchovy spine is 320~\textmu m.}
    \label{fig:xray-images}
\end{figure}

To demonstrate feasibility of high-resolution imaging using the DCCS architecture, a set of test objects was imaged at 50~keV (Figure \ref{fig:xray-images}).
A bean pod, a dried anchovy, and a chili pepper were acquired at a local grocery store, bound with adhesive tape, and mounted onto the x-ray sample stage.
A photograph of the bound objects is shown in Figure \ref{fig:xray-images}(a).
A composite image produced with $3.2 \times 10^{10}$ x-ray photons is shown with a vignette to isolate the region of interest in Figure \ref{fig:xray-images}(b) with an inset scale bar.
Considering a source-to-sample distance of 1.55~m and a source-to-detector distance of 5.66~m, there is a geometric magnification factor of 3.65.
This results in an effective detector pixel size of 20.5~\textmu m.
The width of the anchovy spine was measured to be 320~\textmu m, and even smaller features are evident in the image.
Since the lower bound on the focused laser spot size is 5~\textmu m, this sets a lower bound on the potential imaging resolution of the laser-Compton x-ray source.
With that in mind, the resolution of the image in \ref{fig:xray-images}(b) is detector-limited.
It is also worth noting that although this imaging set-up was not optimized for phase-based imaging applications, there is evidence of edge enhancement in Figure \ref{fig:xray-images}(b).
Edge enhancement effects are most obvious around the seeds inside of the chili pepper.

\section{FLASH-Relevant Electron Beam Results}
\label{sec:vhee-results}

To demonstrate the feasibility of operating the DCCS architecture with 1000 electron microbunches per macrobunch at 100~Hz, temporary modifications were made to the photogun laser system.
Using the pulse synthesis scheme in Figure \ref{fig:pulse-synthesis}(a), the bit pattern generator was programmed to produce trains of 1000 microbunches at 800~kHz and subsequent amplification and repetition rate reduction steps were adjusted accordingly.
Although the currently operating DCCS prototype at Lumitron Technologies is not fully configured to optimally run at 1000 microbunches, this adjustment was made for an initial demonstration study.
The multi-pass cell typically used to spectrally broaden the photogun laser (PGL) pulses before final compression \cite{seggebruch:2024} was bypassed to maximize micropulse energy before frequency conversion to the UV.
Without extra spectral broadening and subsequent compression, the amplified PGL pulse width increased to about 14 ps, which was not ideal for low-emittance electron beam operation.
For this initial demonstration of high-charge electron beam operation, the goal was to evaluate the feasibility of 1000-bunch operation at high charge and moderate electron energy, to understand beam stability under those operating conditions, and to evaluate the structural integrity of a custom diamond exit window assembly for electron irradiation.

\begin{figure}[ht!]
    \centering
    \includegraphics[width=13cm]{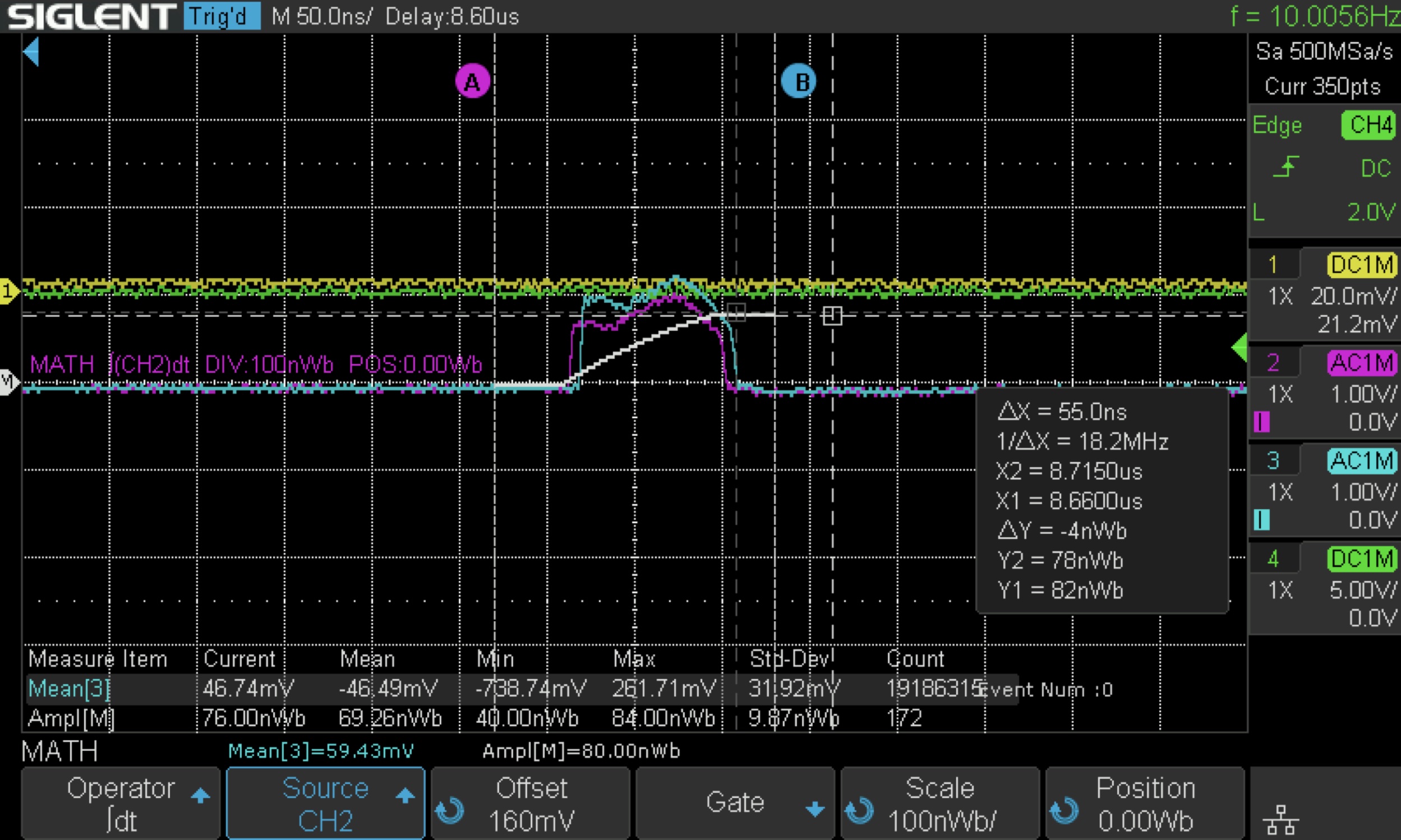}
    \caption{Screenshot of two ICT traces taken on a Siglent oscilloscope. The pulse widths of the ICT traces are consistent with the expected 86.6~ns length of a train of 1000 electron microbunches spaced at 87.5~ps. CH3 (cyan) is the ICT signal just after the photogun while CH2 (magenta) is the signal from the ICT after all accelerator sections just before the laser-electron interaction chamber. An integration of CH2 (white) is carried out to calculate the total charge in the macrobunch. From the ``Ampl[M]'' Measure Item on the bottom of the screen, an integrated signal of $69.62 \pm 9.87$ nWb was measured over 172 macrobunches. This corresponds to a macrobunch charge of $13.85 \pm 1.97$ nC.}
    \label{fig:ICT-trace}
\end{figure}

The electron beam was successfully accelerated to 49.4~MeV using Lumitron's built-in-house X-band photogun and three T53VG3 sections operating at 100~Hz.
The photogun and each accelerating structure were fed RF power from separate klystron/modulator units each capable of operation up to 400~Hz.
For this demonstration, all RF power was delivered at 100~Hz.
The first section received 10~MW of peak RF power, the second accelerator section received 5.7~MW of peak RF power, and the third accelerator section received 19~MW of peak RF power.
The RF power to the sections was tuned so that the mean electron energy would be near 50~MeV as measured by a 35$^{\circ}$ dipole magnet spectrometer available in the beamline.
Beam quality degradation was noted with a 1\% tail on the electron beam spectrum that otherwise reflected a symmetric energy spread of 0.2\%.
Figure \ref{fig:ICT-trace} illustrates the measured traces from two separate Bergoz integrating current transformers (ICTs) with custom 5-ns output pulse widths. The total charge in each electron macrobunch could be measured just after the photogun (cyan) and just after the laser-electron interaction chamber (purple) near the vacuum exit window.
More charge loss than usual was noted through the accelerator line, most likely because of the large electron energy spread and deteriorated beam quality due to the relatively long PGL micropulses.
After accumulating statistics over 172 electron macrobunches, an integrated signal (white) over the downstream ICT trace (magenta) measured $69.62 \pm 9.87$ nWb as a typical macrobunch signal.
Dividing this by the manufacturer-reported ICT conversation factor of 5 V/A yields a measured macrobunch charge of $13.85 \pm 1.97$ nC.

\begin{figure}[ht!]
    \centering
    \includegraphics[width=15cm]{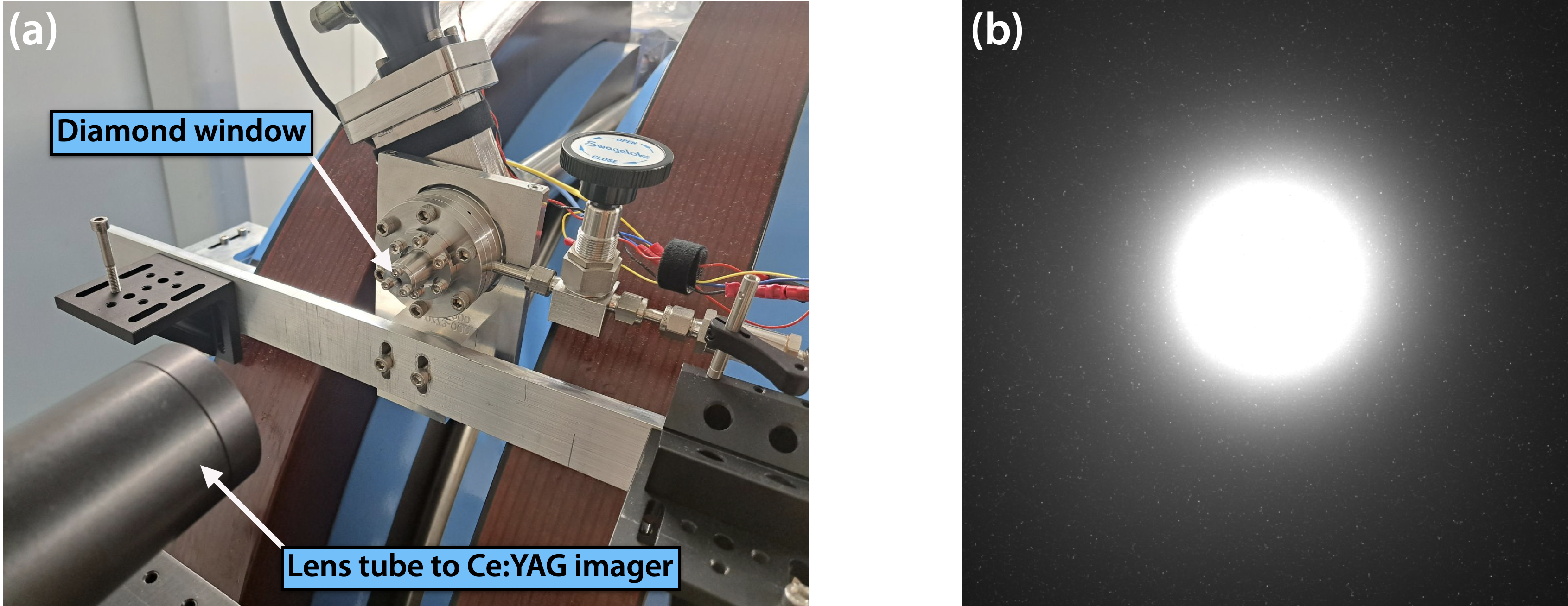}
    \caption{(a) Installed diamond window assembly at the end of vacuum beamline is pictured in center. In the bottom left foreground is a lens tube holding a Ce:YAG scintillator screen that is imaged onto a camera sensor. (b) Measured Ce:YAG scintillation signal indicating a highly collimated output from the diamond exit window and successful transport of electrons into air.}
    \label{fig:vhee-setup}
\end{figure}

Figure \ref{fig:vhee-setup}(a) shows the 100-\textmu m-thick mounted diamond exit window assembly at the end of the beam pipe that was used during this electron beam demonstration.
Larger aperture window assemblies optimized for both x-ray and electron beam operation are currently being developed for future experiments.
A Ce:YAG imaging system was placed just downstream of the exit window to measure the scintillation signal of the electron beam.
When the electron beam was fully aligned to the window, the camera was saturated when set to operate with an exposure time of 10~ms to capture the contributions of a single macrobunch.
Although the image in \ref{fig:vhee-setup}(b) cannot provide a quantitative measure of electrons leaving vacuum, it confirms successful operation of the diamond exit window assembly and the presence of a collimated ionizing beam.
An in-air ICT is currently being deployed just downstream of the exit window to directly measure the electron charge.

While these initial electron beam results were performed at 49.4~MeV, the DCCS accelerator has produced up to 70~MeV electron beams to date, and is configured to produce 100~MeV in its current state.
Upgrades are planned for 137~MeV operation in the next 12~months from the time of writing this manuscript for advanced high-energy imaging applications of relevance to industrial nondestructive tests.
The current 100~MeV setup is sufficient for \textit{in vivo} animal irradiation studies planned in the near term.
While 14~nC was initially generated per macrobunch in this demonstration, upwards of 50~nC per macrobunch are readily possible by extending the bunch train to 3000~microbunches.
Extending bunch train length further is limited by the temporal length of the compressed RF pulses that drive the accelerator sections.

\section{A Vision for Image-Guided VHEE Radiation Therapy}
\label{sec:image-guided}

The concomitant, inherently colinear production of high-brightness x rays through laser-Compton scattering and VHEEs provides a unique opportunity for in-line image-guided radiation therapy using a single device.
The notion that a Compton-scattered x-ray beam could be used for image guidance for an underlying VHEE source was, to the best of our knowledge, first suggested by DesRosiers, \textit{et al.} \cite{desrosiers:2008} in the context of laser-plasma VHEE sources.
Although DesRosiers, \textit{et al.} (2008) \cite{desrosiers:2008} pointed to the potential benefits in x-ray quality, comparing laser-Compton x rays to those produced by synchrotron facilities, there have not been, to the best of our knowledge, any discussions on the feasibility of a laser-Compton x-ray image-guided VHEE radiation therapy system.
In this section, we aim to outline the challenges and propose solutions for implementing such a system.

Fundamentally, this concept requires the rapid switching between the production of laser-Compton x rays to the delivery of a VHEE beam.
The most straightforward implementation of this switching, considering the DCCS architecture, would be to shutter both ILS and PGL lasers, turn off the final electron bending magnet, and then release bursts of PGL pulses based on the prescribed radiation treatment.
In this operation mode, the minimum time between imaging and therapy is the time required to operate laser shutters and the decay time of the bending magnet, which may be on the order of seconds.
For this approach to work, the electron energy used to produce LCS x rays must also be of an energy relevant for the radiation treatment.
For the DCCS architecture presented here, producing a 70~keV imaging beam, which has been previously identified as an energy of interest for minimizing dose during phase contrast imaging applications \cite{reutershan:2022}, requires 45~MeV electrons.
Recent treatment planning simulation studies have suggested that this energy range may be of interest, especially when operated at FLASH-relevant dose rates, for treating pediatric brain tumors \cite{breitkreutz:2020}.
However, 45~MeV electrons are insufficient for use in deep-seated tumors in adult humans, and simply increasing the electron energy will correspond with an x-ray beam whose energy is too high for practical clinical imaging applications.
One of the appeals of producing x rays or $\gamma$ rays through laser-Compton scattering is to minimize the energy requirements on the electron beam, thus enabling the use of compact accelerator architectures, especially when compared to the required synchrotron facility requirements to produce the same x-ray or $\gamma$-ray energy.
Additionally, counter-propagating electron and laser beams uses the \textit{least} energetic electrons possible for a given scattered x-ray energy.

If one could quickly adjust the electron beam energy after x-ray imaging, then the combinations of imaging x-ray energies and therapeutic electron beam energies increases dramatically.
In this operation mode on the DCCS, imaging could be performed using an optimal laser-Compton spectrum for either minimizing dose, maximizing contrast, or any combination of metrics for the specific imaging task.
Then, after tuning the electron beam energy by changing the amount of RF power delivered to the accelerating sections, an electron beam could be delivered for image-guided radiation therapy.
Activities at Lumitron have demonstrated the ability to manually tune the system from one energy to the next with better than 1\% accuracy in less than 10 minutes without relying on any preset values for RF power and steering magnets.
In principle, with appropriate preset values established, electron beam energy can be adjusted on the timescale of seconds.

Another consideration for image-guided VHEE FLASH radiation therapy is identifying appropriate combinations of x-ray and electron beam transverse sizes for imaging and therapy respectively.
For example, the 38-MeV electron beam considered in the manuscript in Section \ref{sec:compton-results} resulted in a usable x-ray imaging field of about 8 cm at a distance of 5.66 m from the laser-electron interaction point.
Although this field-of-view is smaller than what is used with current clinical x-ray sources, a benefit of using a low-divergence beam is that the detector can be placed further away from the patient while preserving the x-ray image.
Scattered x rays from the patient are naturally filtered away, improving the image quality on the detector.
Additionally, placing the detector further away from the patient enables phase-based image techniques, which can improve the differentiation of soft tissue for increased diagnostic potential.
For the electron beam, preclinical small-animal studies are a primary short-term objective to demonstrate the feasibility of UHDR electron beam irradiation using the DCCS.
Beams of this type can be constructed using natural beam expansion from the exit window and simple collimators and scattering foils \cite{bourgouin:2022}.
For clinical applications, VHEE pencil-beam scanning is a potential modality, and a proposed magnet kicker system has recently been discussed \cite{stephan:2022}.
However, pencil-beam scanning would result in an ``image-informed'' as opposed to a strictly ``image-guided'' modality in which the laser-Compton x-ray image can define the extent of the scan, but is no longer inherently colinear with the electron beam since the electron beam path is modified by a steering magnet.
In general, as long as any electron beam shaping does not modify the central trajectory of the electron beam, the potential for x-ray image guidance is preserved.

Previously presented work has investigated the feasibility of using the free propagating x-ray and electron beams after passing through a common vacuum exit window, and has identified operating conditions of 37~MeV electrons and 51~keV Compton-scattered x rays to potentially image and subsequently irradiate a target the size of a mouse skull \cite{effarah:2024}.
51~keV was chosen as the imaging energy to utilize a Gd foil to produce a K-edge hole that can be used as an image-guidance crosshair to more accurately identify the central propagation axis of the electron beam (see Figure \ref{fig:gd-k-hole}).
To fully explore the parameter space available, the incorporation of electron beam optics, x-ray optics, and considerations of different laser-electron interaction geometries should be considered, and will be explored further in future work.

\section{Conclusion}

Very high energy electrons (VHEEs) have been identified as a promising ionizing radiation modality, especially in the context of eliciting the FLASH effect, but VHEE source size and access to high-fidelity image guidance limit clinical implementation.
In this work, we described the Distributed Charge Compton Source (DCCS) as a uniquely suited architecture for image-guided VHEE FLASH radiation therapy.
Through maximally distributing electron charge in long, low-charge trains of microbunches at 11.424~GHz (X-band), large currents can be accelerated without compromising the quality of the electron beam.
This conservation of electron beam quality enables the electron beam to be used not only directly as an ionizing radiation source, but also as a generator of secondary x rays through the head-on collision with a counter-propagating train of laser pulses through the process of laser-Compton scattering.
Since the laser micropulses and electron microbunches are generated using the same RF master lock, the pulse/bunch trains are inherently synchronized up to a simple phase delay.
As a systems integration test, the DCCS has demonstrated the production of laser-Compton x rays with narrow energy bandwidths (54~keV with 0.41\% on-axis RMS bandwidth presented here) and the production of electron macrobunches with FLASH-relevant charge densities (14~nC in 86.6~ns at 100~Hz).
As the underlying systems are continued to be commissioned, a primary milestone is to use the same electron beam for both the production of inherently colinear laser-Compton x rays and electron irradiation in a single experimental session, thus demonstrating the feasibility of using the DCCS architecture for x-ray image-guided VHEE FLASH radiation therapy.

\section*{Ethics statement}
Ethical approval was not required for the study involving animals in accordance with the local legislation and institutional requirements because the animal product (dried anchovy) used for imaging tests was purchased at a grocery store.

\section*{Conflict of Interest Statement}

CPJB, JMA, AJA, JCRB, SMB, MMC, RADLL, DAD, HHE, RF, AG, KJG, ASG, FVH, YH, GI, MJ, CAJ, KWK, AL, RJL, MWM, EM, CLN, HJN, KRP, ZRP, MEQ, FR, TR, JS, MES, MWLS, JYY, CBZ, LEZ, and EJZ were employed by Lumitron Technologies, Inc. 
CPJB holds patents related to the underlying research presented in this work.

\section*{Author Contributions}

Conceptualization: CPJB, JMA, HHE, FVH, YH, GI, AL, FR, TR, LEZ;
Funding Acquisition, CPJB;
Project administration: CPJB, JMA, RF, KJG, FVH, GI, FR, LEZ;
Machine Fabrication, Assembly, and Test: CPJB, JMA, AJA, JCRB, SMB, MAC, MMC, MED, RADLL, DAD, HHE, RF, AG, KJG, ASG, FVH, LH, YH, GI, MJ, CAJ, KWK, AL, RJL, MWM, EM, CLN, HJN, KRP, ZRP, MEQ, FR, KR, TR, JS, MES, MWLS, NHY, CBZ, LEZ, EJZ, JZ;
Investigation: CPJB, AJA, JCRB, SMB, MMC, RADLL, HHE, AG, FVH, LH,  YH, GI, MJ, AL, RJL, EM, KRP, MEQ, FR, KR, TR, MES, MWLS, LEZ, JZ;
Formal Analysis: HHE, YH, AL, TR;
Software: JMA, AJA, JCRB, HHE, FVH, LH, YH, AL, MEQ, JS, TR;
Visualization: CPJB, JMA, SMB, HHE, YH, AL, MWM, KRP, FR, ZRP, TR, JS, MWLS, JYY, NHY;
Writing - Original Draft: CPJB, HHE;
Writing - Review \& Editing: CPJB, JMA, AJA, JCRB, SMB, MAC, MMC, MED, RADLL, DAD, HHE, RF, AG, KJG, ASG, FVH, LH, YH, GI, MJ, CAJ, KWK, AL, RJL, MWM, EM, CLN, HJN, KRP, ZRP, MEQ, FR, KR, TR, JS, MES, MWLS, JYY, NHY, CBZ, LEZ, EJZ, JZ;
Supervision: CPJB

\section*{Funding}
This work was supported by funds from Lumitron Technologies, Inc and in part by the DARPA GRIT program under Contract Number HR00112090059. HHE and TR received partial training funding through the NIH T32GM008620.

\section*{Acknowledgments}
The authors wish to acknowledge the following individuals for their varied contributions to the establishment of prior configurations of the hardware described in this document and/or the creation of relevant machine-related infrastructure:
Ted Andreas, Lisset Ayala, Sam	Chu, Mackenzie E. Drobny, Jake Ewing, Derek Fietz, Christopher Fluxa, Daniel Gitlin, Albert Herrero Parareda, Jason Holtkamp, Salma Idrus, Patrick Lancuba, Mealaud Mokhtarzad, Eric C. Nelson, Christine V. Nguyen, Milan N. Nguyen, Colin Sledge, Ahmed Srass, Vy Tran, Alexandra Valanis, Gabriella Vass, Kinga Vassne, Daniel Vega, and Li Yi-Ning.

\section*{Data Availability Statement}
The data underlying this presented work are not publicly available but can be provided upon reasonable request.

\bibliographystyle{Frontiers-Vancouver} 
\bibliography{bibliography}


\end{document}